%% file: ms.tex
\documentclass{vldb}
\usepackage{balance}  %

\usepackage[]{graphicx}
\usepackage{verbatim}
\usepackage{color}
\usepackage{cite}
\usepackage{subfig}
\usepackage[font={small}]{caption}
\usepackage[binary-units]{siunitx}
\usepackage{abbrevs}
\usepackage{pifont}
\usepackage{xspace} %
\usepackage{listings}
\usepackage[normalem]{ulem}
\usepackage{enumerate}
\usepackage{booktabs}
\usepackage[export]{adjustbox}
\usepackage[inline]{enumitem}

\definecolor{darkgreen}{rgb}{0.00,0.39,0.00}
\usepackage[hyphens]{url}
\usepackage[
    colorlinks,
    linkcolor=blue,
    citecolor=darkgreen,
    urlcolor=blue,
]{hyperref}

\newcommand{\mypar}[1]{\vspace{-9pt}~\\\noindent{\bf #1}}

\newcommand{\evidence}{\vspace{-7pt}~\\\noindent{\textbf{Discussion. }}}
\newcommand{\guideline}{\vspace{-7pt}~\\\noindent{\textbf{Guidelines. }}}
\newcommand{\pitfallN}[3]{\noindent{\textbf{Pitfall #1: #2. }\textit{#3}}}
\newcommand{\pts}{PTS}
\newcommand{\ptses}{PTSes}

\definecolor{red3}{rgb}{0.80,0.00,0.00}
\definecolor{redx}{rgb}{0.50,0.00,0.00}

\newcommand{\remove}[1]{}

\makeatletter
\let\maybe@space@\xspace
\makeatother

\newabbrev\SCM{storage capacity manager (SCM)}[SCM]
\newabbrev\GC{garbage collection (GC)}[GC]
\newabbrev\SMR{Shingled Magnetic Recording (SMR)}[SMR]
\newabbrev\SSD{Solid State Drives (SSDs)}[SSD]

\newcommand{\CM}[1]{}

\DeclareSIUnit{\cycles}{cycles}

\DeclareSIUnit{\seconds}{secs}

\newcommand{\RDB}{RocksDB\xspace}
\newcommand{\WT}{WireTiger\xspace}

\newcommand{\wad}{WA-D}
\newcommand{\wau}{WA-A}

\vldbTitle{Toward a Better Understanding and Evaluation of Tree Structures on Flash SSDs [EA\&B]}

\begin{document}
\makeatletter
\def\@copyrightspace{\relax}
\makeatother

\date{}

\title{Toward a Better Understanding and Evaluation\\ of Tree Structures on Flash SSDs}

\numberofauthors{3}
\author{
\alignauthor
Diego Didona\\
\affaddr{IBM Research Europe}\\
\textit{ddi@zurich.ibm.com}
\alignauthor
Nikolas Ioannou\\
       \affaddr{IBM Research Europe}\\
\textit{nio@zurich.ibm.com}\\
\alignauthor Radu Stoica\\
       \affaddr{IBM Research Europe}\\
\textit{rst@zurich.ibm.com}\\
\and  %
\alignauthor Kornilios Kourtis\titlenote{Work done while at IBM Research.}\\
       \affaddr{Independent researcher}
\textit{kkourt@kkourt.io}\\
}
\maketitle

\sloppy

\input{abstract}
\input{introduction}

\input{background}
\input{methodology}
\input{evaluation}
\input{related}
\input{conclusion}
\balance
\bibliographystyle{abbrv}
\bibliography{ms}

\end{document}

%% file: abstract.tex
\begin{abstract}
Solid-state drives (SSDs) are extensively used to deploy persistent data stores, as they provide low latency random access, high write throughput, high data density, and low cost. Tree-based data structures are widely used to build persistent data stores, and indeed they lie at the backbone of many of the data management systems used in production and research today. 

In this paper, we show that benchmarking a persistent tree-based data structure on an SSD is a complex process, which may easily incur subtle pitfalls that can lead to an inaccurate performance assessment. At a high-level, these pitfalls stem from the interaction of complex software running on complex hardware.
On one hand, tree structures implement internal operations that have nontrivial effects on performance. On the other hand, SSDs employ firmware logic to deal with the idiosyncrasies of the underlying flash memory, which are well known to lead to complex performance dynamics.

We identify seven benchmarking pitfalls using RocksDB and WiredTiger, two widespread implementations of an LSM-Tree and a B+Tree, respectively. We show that such pitfalls can lead to incorrect measurements of key performance indicators, hinder the reproducibility and the representativeness of the results, and lead to suboptimal deployments in production environments. 
We also provide guidelines on how to avoid these pitfalls to obtain more reliable performance measurements, and to perform more thorough and fair comparison among different design points.
~\\
\end{abstract}

%% file: introduction.tex
\section{Introduction}
\label{sec:introduction}
Flash solid-state drives (SSDs) are widely used to deploy persistent data storage in data centers, both in bare-metal and cloud deployments~\cite{mcdipper,fatcache,rocksdb,leveldb,ssd-dc1,ssd-dc2,schroeder:fast:2016}, while also being an integral part of public cloud offerings~\cite{ibm:ssd,aws:ssd,google:ssd}.
While SSDs with novel technologies such as 3D Xpoint~\cite{micron,optane} offer significant advantages~\cite{lepers:sosp:2019,kourtis:fast:2019,wu:damon:2019}, they are not expected to replace flash-based SSDs anytime soon. These newer technologies are not yet as mature from a technology density standpoint, and have a high cost per GB, which  significantly hinders their adoption. Hence, flash SSDs are expected to be the storage medium of choice for many applications in the short and medium term future~\cite{future}. 

Persistent tree data structures (\ptses{}) are widely used to index large datasets. \ptses{} are particularly appealing, because, as opposed to hash-based structures, they allow for storing the data in sorted order, thus enabling efficient implementations of range queries and iterators over the dataset.  Examples of \ptses{} are the log structured merge (LSM) tree~\cite{oneil:lsm:1996}, used, e.g., by RocksDB~\cite{rocksdb} and BigTable~\cite{chang:tocs:2008}; the B+Tree~\cite{comer:csur:1979}, used, e.g., by Db2~\cite{db2} and WiredTiger~\cite{wiredtiger} (which is the default storage engine in MongoDB~\cite{mongodb}); the Bw-tree~\cite{Levandoski:sigmod:2013}, used, e.g., by Hekaton~\cite{diaconu:sigmod:2013} and Peloton~\cite{pavlo:cidr:2017,peloton}; the B-$\epsilon$ tree~\cite{brodal:soda:2003}, used, e.g., by Tucana~\cite{Papagiannis:atc:2016} and TokuMX~\cite{tokumx}.

Over the last decade, due to the reduction in the prices of flash memory, \ptses{} have been increasingly deployed over flash SSDs~\cite{price1,price2}. Not only  do \ptses{} use flash SSDs  as a drop-in replacement for hard disk drives,  but new \pts{} designs are specifically tailored to exploit the capabilities of flash SSDs and their internal architectures~\cite{wang:bwtree:sigmod18,wang:eurosys:2014,lu:tos:2017,shen:tos:2018,trivedi:tos:2018}. 

\mypar{Benchmarking \ptses{} on flash SSDs.} Given their ubiquity, evaluating accurately and fairly the performance of \ptses{} on flash SSDs is a task of paramount importance for industry and research alike, to compare alternative designs. Unfortunately, as we show in this paper, such task is a complex process, which may easily incur subtle pitfalls that can lead to an inaccurate and non-reproducible performance assessment.  

The reason for this complexity is the intertwined effects of the internal dynamics of flash SSDs and of the \pts{} implementations.
On the one hand, flash SSDs employ firmware logic to deal with the idiosyncrasies of the underlying flash memory, which result in highly non-linear dynamics~\cite{Hu:systor:2009,stoica:vldb2013,desnoyer:tocs:2014,stoica:mascots:2019}.
On the other hand, \ptses{} implement complex operations, (e.g., compactions in LSM-Trees and rebalancing in B+Trees), and update policies (e.g.,  in a log-structured fashion vs in-place). These design choices are known to lead to performance that are hard to analyze~\cite{lsm-bush:sigmod19}. They also result in widely different access patterns towards the underlying SSDs, thus leading to complex, intertwined performance dynamics.

\mypar{Contributions.} In this paper, we identify seven benchmarking pitfalls which relate to different aspects of the evaluation of a \pts{} deployed on a flash SSD.  Furthermore, we provide guidelines on how to avoid these pitfalls. We provide specific suggestions both to academic researchers, to improve the fairness, completeness and reproducibility of their results, and to performance engineers,  to help them identify the most efficient and cost-effective \pts{} for their workload and deployment.

In brief, the pitfalls we describe and their consequences on the \pts{} benchmarking process are the following:

\begin{enumerate}[label={\bf \arabic*.}]
\itemsep0em 
    \item {\bf Running short tests.} Flash SSDs have time-evolving performance dynamics. Short tests lead to results that are \emph{not} representative of the long-term application performance.

    \item {\bf Ignoring device write amplification (WA-D).} SSDs perform internal garbage collection that leads to write amplification. Ignoring this metric, leads to inaccurate measurements of the I/O efficiency of a \pts{}.

    \item {\bf Ignoring internal state of the SSD.} Experimental results may significantly vary depending on the initial state of the drive. This pitfall leads to unfair and non-reproducible performance measurements.

    \item {\bf Ignoring the effect of the dataset size on SSD performance.} SSDs will exhibit different performance depending on the amount of data stored. This pitfall leads to biased evaluations.

    \item {\bf Ignoring the extra storage capacity that a \pts{} needs.} This pitfall leads to ignore the storage versus performance trade-off of a \pts{}, which is methodologically wrong and can result in sub-optimal deployments in production systems.

    \item {\bf Ignoring software overprovisioning.} Performance of SSDs can be improved by overprovisioning storage space. This pitfall leads to ignore the storage versus performance trade-off achievable by a \pts{}.

     \item {\bf Ignoring the effect of the underlying storage technology on performance.} This pitfall leads to drawing quantitative and qualitative conclusions that do not hold across different SSD types.

\end{enumerate}

We present experimental evidence of these pitfalls using an LSM-Tree and an B+Tree, the two most widely used \ptses{}, which are also at the basis of several other \ptses{} designs~\cite{bortnikov:vldb:2018,ren:vldb:2017,luo:vldb:2020}. In particular, we consider the LSM-Tree implementation of RocksDB, and the B+Tree implementation of WiredTiger. We use these two systems since they are widely adopted in production systems and research studies.

The storage community has studied the performance of flash SSDs extensively, focusing on understanding, modeling, and benchmarking their performance~\cite{stoica:vldb2013,desnoyer:tocs:2014,tavakkol:fast:2018,ioannou:mascots:2018}. Despite this, we have found that many works from the  systems and databases communities are not aware of the pitfalls in evaluating \ptses{} on SSDs. Our work offers a list of factors to consider, and bridges this gap between these communities by illustrating the intertwined effects of the complex dynamics of \ptses{} and flash SSDs. Ultimately, we hope that our work paves the way for a more rigorous, fair, and reproducible benchmarking of \ptses{} on flash SSDs.

\mypar{Outline.} The remainder of the paper is organized as follows. Section~\ref{sec:background} provides background on the internals of flash SSDs, LSM-Trees and B+Trees. Section~\ref{sec:testbed} describes the experimental setup of our study. Section~\ref{sec:pitfalls} presents the results of our experimental analysis and discusses the benchmarking pitfalls that we identify. Section~\ref{sec:rw} discusses related work. Section~\ref{sec:conclusion} concludes the paper.

\remove{

\begin{enumerate}
\item {\em Not reaching steady state performance.} SSDs implement complex firmware logic, which has highly nonlinear effects on performance, and whose internal dynamics settle after a potentially long transitory phase. We show that failing to take into account this transitory phase can lead to wrong performance measurements.

\item {\em Not analyzing I/O amplification at the SSD level.} SSDs lack support for in-place updates, and hence employ complex garbage collection logics to implement a log-structured storage. We show that ignoring the additional data writes generated by garbage collection leads to erroneous conclusions on the I/O efficiency and flash-friendliness of a tree data structure. 

\item {\em Not controlling the initial state of the SSD.} The steady-state performance of an SSD may depend highly on the initial state of the SSD, in terms of written/clean blocks. We show that performing benchmarks from different SSD initial conditions can lead both to an incorrect analysis of the performance of a data structure, and to a wrong comparison among data structures.
\end{enumerate}
}

%% file: background.tex
~\\
\section{Background}
\label{sec:background}
\begin{figure}
\includegraphics[scale=0.45]{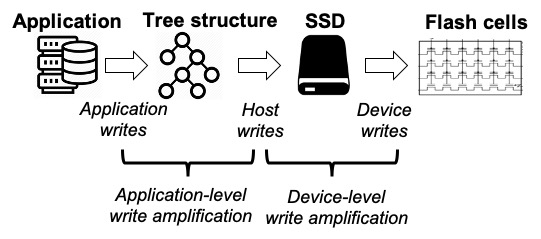}
\caption{Write flow and amplification in \ptses{}. The \pts{} issues additional writes to the SSD to maintain its internal tree structure, which lead to application-level write-amplification. The SSD firmware performs additional writes to overcome the lack of in-place update capabilities of flash memory, which lead to device-level write amplification.}
\label{fig:writes}
\end{figure}
Section~\ref{sec:background:kv} provides on overview of the \ptses{} that we use to demonstrate the benchmarking pitfalls, namely LSM-Trees and B+Trees. 
Section~\ref{sec:background:ssd} provides background on the key characteristics of flash-based SSDs that are related to the benchmarking pitfalls we describe. 
Figure~\ref{fig:writes} shows the flow of write operations from an application deployed over a \pts{} to the flash memory of an SSD.
\subsection{Persistent Tree Data Structures}
\label{sec:background:kv}
We start by introducing the two \ptses{} that we use in our experimental study, and two key metrics of the design of a  \pts.
\subsubsection{LSM-Trees}
LSM-Trees~\cite{oneil:lsm:1996} have two main components: an in-memory component, called {\em memtable}, and a disk component. Incoming writes are buffered in the memtable. Upon reaching its maximum size, the memtable is flushed onto the disk component, and a new, empty memtable is set up. The disk component is organized in levels $L_1,\cdots,L_N$, with progressively increasing sizes. $L_1$ stores the memtables that have been flushed. Each level $L_i, i>1$, organizes data in sorted files that store disjoint key ranges. When $L_i$ is full, part of its data is pushed to $L_{i+1}$ through an operation called {\em compaction}. Compaction merges data from one level to the next, and discards older versions of duplicate keys, to maintain disjoint key ranges in  each level.
\subsubsection{B+Trees}
B+Trees~\cite{comer:csur:1979} are composed of internal nodes and leaf nodes. Leaf nodes store key-value data. Internal nodes contain information needed to route a request for a  target key to the corresponding leaf.  Writing a key-value pair entails writing to the appropriate leaf node. A key-value write may also involve modifying the internal nodes to update routing information, or to perform re-balancing of the tree.
\subsubsection{Application-level write amplification}
Internal operations such as flushing and compactions in LSM-Trees, and internal node updating in B+Trees incur extra writes to persistent storage. These extra writes are detrimental for performance because they compete with key-value write operations for the SSD bandwidth, resulting in lower overall throughput and higher latencies~\cite{sears:sigmod:2012,balmau:atc:2017,lepers:sosp:2019,luo:vldb:2020}. 
We define application-level write amplification (\wau{}) the ratio between the overall data written by the \pts{} (which considers  both application data and internal operations) and the amount of application data written. \wau{} is depicted in the left part of Figure~\ref{fig:writes}.

\subsubsection{Space amplification}
A \pts{} may require additional capacity of the drive, other than the one needed to store the latest value associated with each key. LSM-Trees, for example, may store multiple values of the same key in different levels (the latest of which is at the lowest level that contains the key). B+Trees store routing information in internal nodes, and may reserve some buffers to implement particular update policies~\cite{Wiredtiger:cow}. {\em Space amplification} captures the amount of extra capacity needed by a \pts{}, and it is defined as the ratio of the amount of bytes that the \pts{} occupies on the drive and the size of the application's key-value dataset.
\subsection{Flash SSDs}
\label{sec:background:ssd}
In this section we describe the internal architecture of flash SSDs, as well as key concepts relevant to their performance dynamics.
\subsubsection{Architecture}
Flash-based SSDs organize data in {\em pages}, which are combined in {\em blocks}. A prominent characteristics of flash memory is that pages do not support in-place updates of the pages. A page needs to be erased before it can be programmed (i.e., set to a new value). The erase operation is performed at the block level, so as to amortize its cost.

Flash translation layers (FTLs)~\cite{ftl:csur} hide such idiosyncrasy of the medium. In general, an FTL performs
writes out-of-place, in a log-structured fashion, and maintains a mapping between
logical and physical addresses. When space runs out, a garbage collection process selects some memory blocks, relocates their valid data, and then erases the blocks. 

In the remainder of the paper, for the sake of brevity, we use the term SSD to refer to a flash SSD.

\subsubsection{Over-provisioning}
Over-provisioning is a key technique to enable garbage collection and to reduce the amount of data that it relocates. Over-provisioning means adding extra capacity  to the SSD to store extra blocks used for garbage collection. The more an SSD is over-provisioned, the lower is the number of valid data that needs to be relocated upon performing garbage collection, and, hence, the higher is the performance of the SSD.
SSDs manufacturers always over-provision SSDs by a certain amount.  The user can further implement a software over-provisioning of the SSD by erasing its blocks and enforcing that a portion of the logical block address (LBA) space is never written. This is achieved by restricting the addresses written by the host application, either programmatically, or by reserving a partition of the disk without ever writing to it.

\subsubsection{Device-level write amplification}
Garbage collection reduces the performance of the SSD as it leads to internal re-writing of data in an SSD. 
We define device-level write amplification (WA-D) as the ratio between the amount of data written to flash memory (including the writes induced by garbage collection) and the amount of host data sent to the device.
WA-D is depicted in the right part of Figure~\ref{fig:writes}.

\remove{
WA-D is a first-order metric that significantly influences SSD latency and throughput \cite{}.
From a bench-marking perspective, several issues make it difficult to accurately account for WA-D.
First, when WA-D reaches \textit{steady-state is unclear}. WA-D takes a long time to stabilize and it also exhibits a significant amount of lag time (i.e., WA-D changes are delayed).
Second, WA-D exhibits \textit{wave-like transition periods} when the workload changes. 
This compounds the difficulty of running a typical experimental scenario where data is first loaded and then the workload of interest starts.
Third, user throughput and latency have a non-linear dependency on WA-D.
Most SSDs have a higher internal than external bandwidth which means that the WA-D impact can be mostly hidden up to a certain point before user-level performance is clearly impacted.
}

%% file: methodology.tex
\section{Experimental Setup}
\label{sec:testbed}
This section describes our experimental setup, which includes the \pts{} systems we benchmark, the hardware on which we deploy them, and the workloads we consider.

\subsection{Systems} We consider two key-value (KV) stores: RocksDB~\cite{rocksdb}, that implements an LSM-Tree, and WiredTiger~\cite{wiredtiger}, that implements a B+Tree. 
Both are mature systems widely used on their own and as building blocks of other data management systems. We configure RocksDB and WiredTiger to use small (10 MB) in-memory page caches and direct I/O, so that the dataset \emph{does not fit} into RAM, and both KV and internal operations are served from secondary storage.

\subsection{Workload} Unless stated otherwise, we use the following workload in our tests. The dataset is composed of 50M KV pairs, with 16 bytes keys and 4000 bytes values. 
 The size of the dataset is $\approx$200 GB,  which represent $\approx$50\% of the capacity of the storage device. 
Before each experiment we ingest all KV pairs in sequential order. We consider a write-only workload, where one user thread updates existing KV pairs according to a uniform random distribution. We focus on a write workload as it is the most challenging to  benchmark accurately, both for the target data structures and for the SSDs. We use a single user thread to avoid the performance dynamics caused by concurrent data accesses.  
We also consider variations of this workload to show that the pitfalls we describe apply  to a broad class of workloads. %

\subsection{Metrics} 
To demonstrate our pitfalls, we analyze several application, system and hardware performance metrics.

\noindent{\em i) KV store throughput}, i.e., the number of operations per second completed by the KV store.

\noindent{\em ii) Device  throughput}, i.e.,  the amount of data written per second to the drive as observed by the OS. The device throughput is often used to measure the capability of a system to utilize the available I/O resources~\cite{lepers:sosp:2019}. We measure device throughput using \texttt{iostat}.  
  
\noindent{\em iii) User-level write amplification}, which we measure by taking the ratio of the device write throughput and the  product of the KV store and the size of a KV pair.  By using the device write throughput,  we  factor in also the write overhead posed by the filesystem. We  assume such overhead to be negligible with respect to the amplification caused by the \pts{} itself.

\noindent{\em iv) Application-level write amplification}, which we measure via SMART attributes of the device.

\noindent{\em v) Space amplification}, which we obtain by taking the ratio of the disk total utilization and the cumulative size of the KV pairs in the dataset. Also this metric factors in the overhead posed by the filesystem, which is negligible with respect to the several GB datasets that we consider.
 
~\\For the sake of readability of the plots, unless stated otherwise, we report 10-minutes average values when plotting the evolution of a performance metric over time.

\subsection{State of the drive} 
\label{sec:testbed:state}
We experiment with two different initial conditions of the internal state of the drive.

\noindent{\em $\bullet$ Trimmed.} All blocks of the device are erased (using the \texttt{blkdiscard} utility). Hence, initial writes are stored directly into free blocks and do not incur additional overhead (no WA-D occurs), while updates after the free blocks are exhausted incur internal garbage collection.
A trimmed device exhibits performance dynamics close (i.e., modulo the wear of the storage medium) to the ones of a mint factory-fresh device. This setting is representative of bare-metal standalone deployments, where a system administrator can reset the state of the drive before deploying the KV store, and the drive is not shared with other applications.

\noindent{\em $\bullet$ Preconditioned.}  The device undergoes a preliminary writing phase so that its internal state resembles the state of a device that has been in use.  To this end, we first write the whole drive sequentially, to ensure that all logical addresses have associated data. Then, we issue random writes for an amount of bytes that is twice the size of the disk, so as to trigger garbage collection. In this setting even the first write operation issued by an application towards any page is effectively an over-write operation. This setting is representative of $i)$ consolidated deployments, e.g., public clouds, where multiple applications share the same physical device,  %
 or $ii)$ standalone deployments with an aged filesystem.

~\\These two configurations represent the two endpoints of the spectrum of the possible initial conditions of the drive, concerning the state of the drive's block. In a real-world deployment the initial conditions of the drive would be somewhere in-between these two endpoints.

\subsection{Hardware} We use a machine equipped with an Intel(R) Xeon(R) CPU E5-2630 v4 @ 2.20GHz (20 physical cores, without hyper-threading) and 126 GB of RAM. The machine runs  Ubuntu 18.04 with a 4.15 generic Linux kernel. The machine's persistent storage device is a 400 GB Intel p3600 enterprise-class SSD~\cite{intel:p3600}. Unless stated otherwise, we setup a single partition on the drive, which takes the whole available space. We  mount an ext4 filesystem configured with the \texttt{nodiscard} parameter~\cite{intel-nvme}. %

%% file: evaluation.tex
\begin{figure*}[!ht]
     \subfloat[RocksDB: KV store and device throughput.\label{fig:eval:ss:rdb:xput}]{
       \includegraphics[scale=0.75]{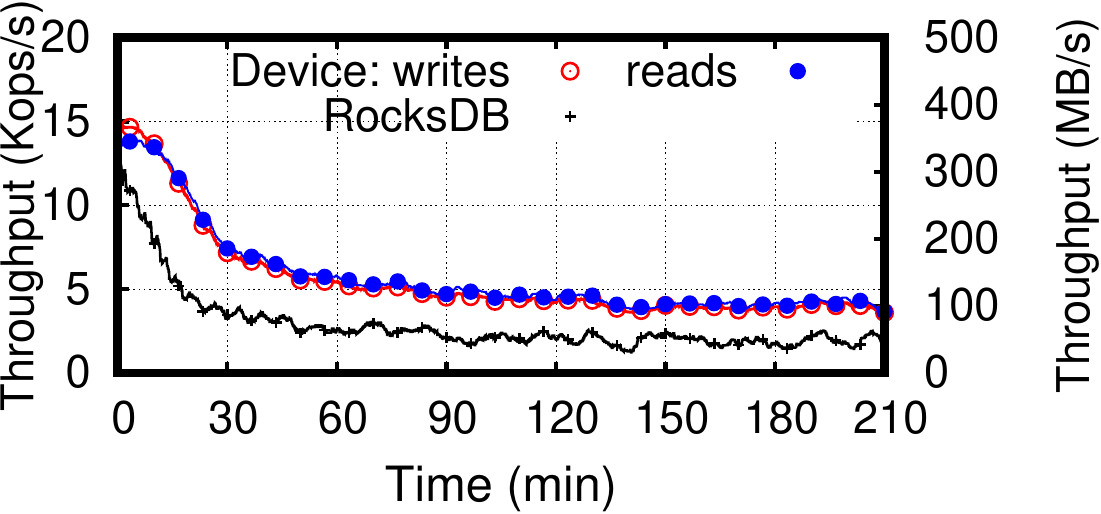}
     }
     \hfill
     \subfloat[WiredTiger: KV store and device throughput.\label{fig:eval:ss:wt:xput}]{
       \includegraphics[scale=0.75]{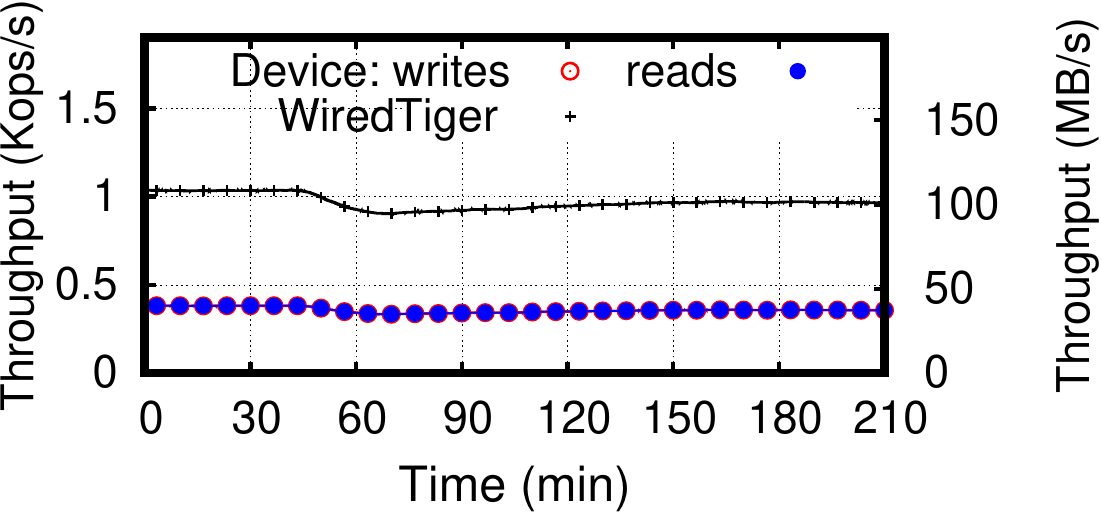}
     }
     \\
       \subfloat[RocksDB: Application and device-level WA.\label{fig:eval:ss:rdb:wa}]{
       \includegraphics[scale=0.75]{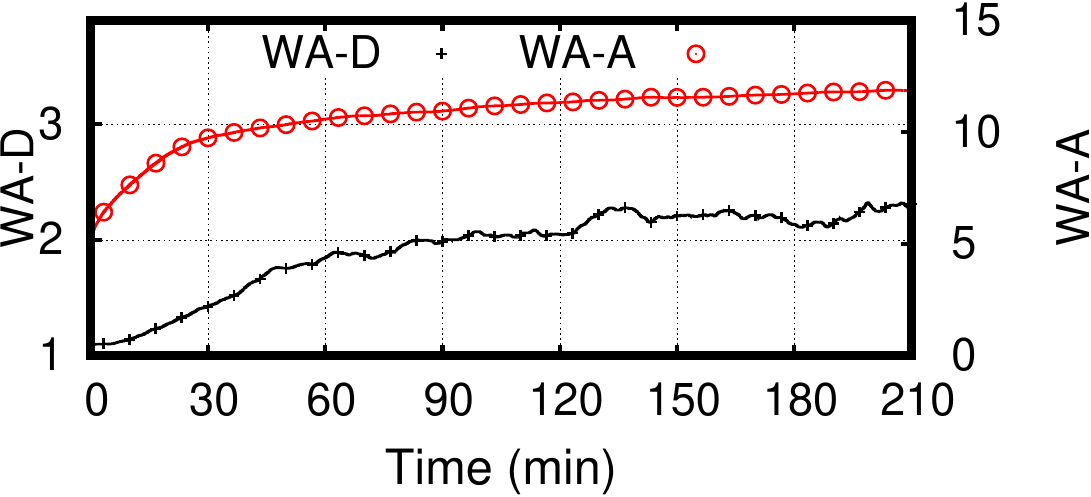}
     } \hfill
     \subfloat[WiredTiger: Application and device-level WA.\label{fig:eval:ss:wt:wa}]{
       \includegraphics[scale=0.75]{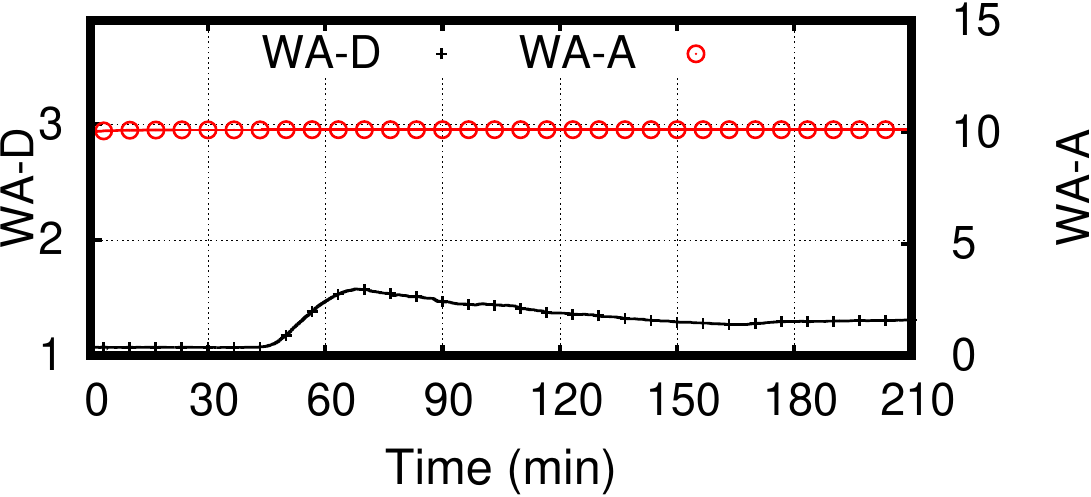}
       }     
     \caption{Difference between steady state and bursty performance (on a trimmed SSD) in RocksDB (left) and WiredTiger (right). Steady-state performance differs from  the initial one (top) because of the change in write amplification (bottom). }\label{fig:eval:ss}
   \end{figure*}
   
\section{Benchmarking Pitfalls}
\label{sec:pitfalls}
This section discuss the benchmarking pitfalls in detail. For each pitfall we  $i)$ first give a brief description; $ii)$ then discuss the pitfall in depth, by providing experimental evidence that demonstrates the pitfall itself and its implications on the performance evaluation process; and $iii)$ finally outline guidelines on how to avoid the pitfall in research studies and production systems.  
\subsection{Steady-state vs. bursty performance} \label{sec:eval:ss}

\pitfallN{1}{Running short tests}%
{Because both PTS and SSD performance varies over time, short-lived tests are unable to capture how the systems will behave under a continuous (non-bursty) workload.}

\evidence{} Figure~\ref{fig:eval:ss} shows the KV store and device throughput (top), and the WA-D and WA-A (bottom) over time, for RocksDB (left) and WiredTiger (right). These results refer to running the two systems on a trimmed SSD. The plots {\em do not} show the performance of the systems during the initial loading phase. The pictures show that, during a performance test, a \pts{} exhibits an initial transient phase during which the dynamics and the performance indicators may widely differ from the ones observed at steady-state. Hence, taking early measurements of a data store's performance may lead to a substantially wrong performance assessment.

Figure~\ref{fig:eval:ss:rdb:xput} shows that measuring the performance of RocksDB in the first 15 minutes would report a throughput of 11-8 KOps/s, which is 3.6-2.6 times higher than the 3 KOps/s that RocksDB is able to sustain at steady-state. In the first 15 minutes, the device throughput of RocksDB is between 375 and 300 MB/s, which is more than 3 times the throughput sustained at steady-state. 

Figure~\ref{fig:eval:ss:rdb:wa} sheds lights on the causes of such performance degradation.  The performance of RocksDB decreases over time for the effect of the increased write amplification, both at the \pts{} and device level.  WA-A increases over time while the levels of the LSM-Tree fills up, and its curve flattens once the layout of the LSM tree has stabilized. WA-D increases over time because of the effect of garbage collection. The initial value of WA-D is close to one, because the SSD is initially trimmed, and keys are ingested in order during the initial data loading, which results in RocksDB issuing sequential writes to the drive. The compaction operations triggered over time, instead, do not result in sequential writes to the SSD flash cells, which ultimately lead to a WA-D slightly higher than 2.

WiredTiger exhibits performance degradation as well, as shown in Figure~\ref{fig:eval:ss:wt:xput}. The performance reduction in  WiredTiger is lower than in RocksDB for three reasons. First, WA-A is stable over time, because updating the B+Tree to accommodate new application writes incurs an amount of extra writes that does not change over time. Second, the increase in WA-D is lower than in RocksDB: WA-D reaches at most the value of 1.7, and converges to 1.5. We analyze more in detail the WA-D of WiredTiger in the next section. Third, WiredTiger is less sensitive to the performance of the underlying device because of synchronization and CPU overheads ~\cite{lepers:sosp:2019}.

\guideline{} Researchers and practitioners should distinguish between steady-state and bursty performance, and prioritize reporting the former. In light of the results portrayed by Figure~\ref{fig:eval:ss}, we advocate that, to detect steady-state behavior, one should implement a holistic approach that encompasses application-level throughput, WA-A, and WA-D.  Techniques such as CUSUM~\cite{cusum} can be used to detect that the values of these metrics do not change significantly for a long enough period of time.

To measure throughput we suggest using an average over long periods of times, e.g., in the order of ten minutes. In fact, it is well known that  \ptses{} are prone to exhibit large performance variations over short period of times~\cite{sears:sigmod:2012,luo:vldb:2019,lepers:sosp:2019}. 
Furthermore, we suggest expressing the WA-A at time $t$ as the ratio of the  {\em cumulative} application writes up to time $t$ and the {\em cumulative} host writes up to time $t$. This is aimed at avoiding oscillations that would be obtained if measuring the WA-D over small time windows. Finally, if WA-D cannot be computed directly from SMART attributes, then we suggest, as a rule of thumb, to consider the SSD as having reached steady-state after the cumulative host writes accrue to at least 3 times the capacity of the drive.  The first device write ensures that the drive is filled once, so that each block has data associated with it. The second write triggers garbage collection, which overwrites the block layout induced by the initial filling of the data-set. The third write ensures that the garbage collection process approaches steady state, and is needed also to account for the (possibly unknown) amount of hardware extra capacity of the SSD (which makes the actual capacity of the drive higher than the nominal capacity exposed to the application).

   \begin{figure*}[!ht]
     \subfloat[RocksDB: throughput.\label{fig:eval:init:rdb:xput}]{
       \includegraphics[scale=0.75]{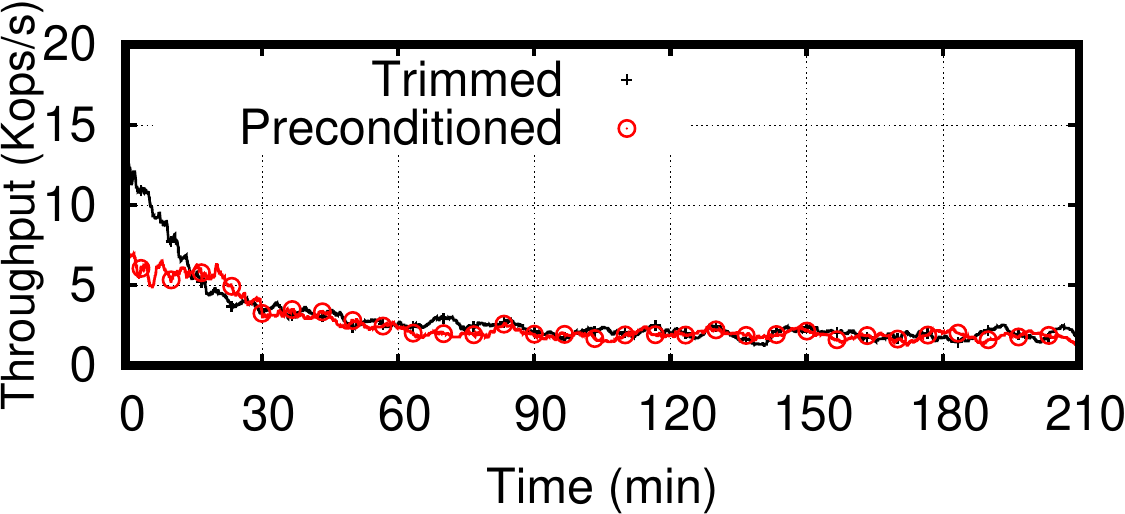}
     }
     \hfill
     \subfloat[WiredTiger: throughput.\label{fig:eval:init:wt:xput}]{
       \includegraphics[scale=0.75]{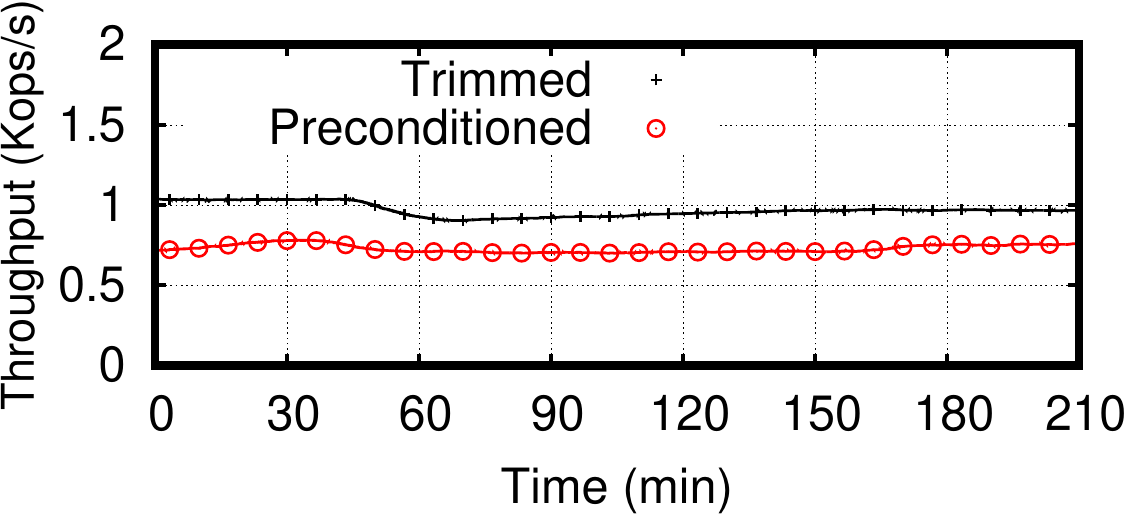}
     }
     \remove{
     \\
     \subfloat[RocksDB device write throughput.\label{fig:eval:init:rdb:ssd}]{
       \includegraphics[scale=0.75]{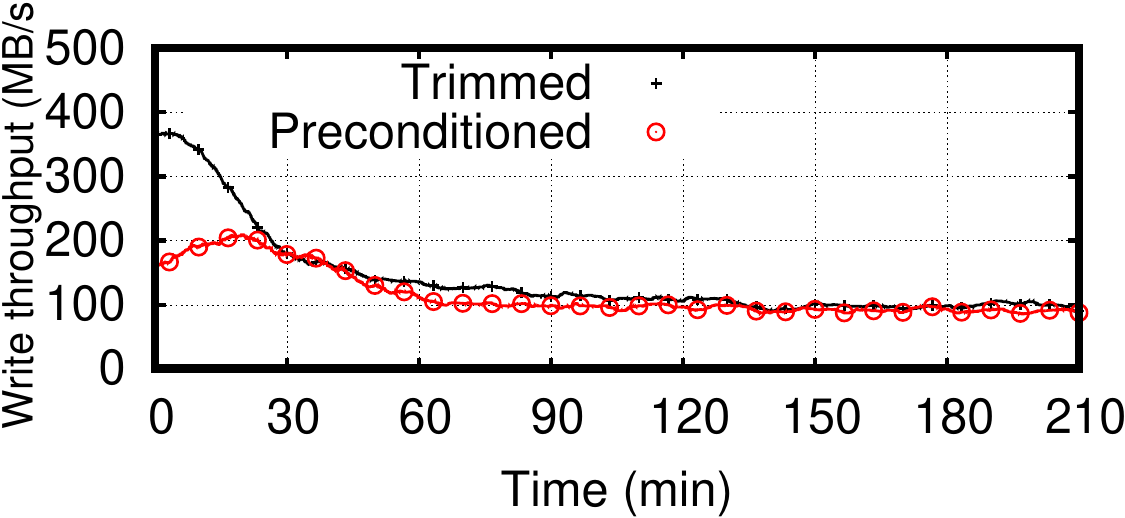}
     }
     \hfill
     \subfloat[WiredTiger: device write throughput.\label{fig:eval:init:wt:ssd}]{
       \includegraphics[scale=0.75]{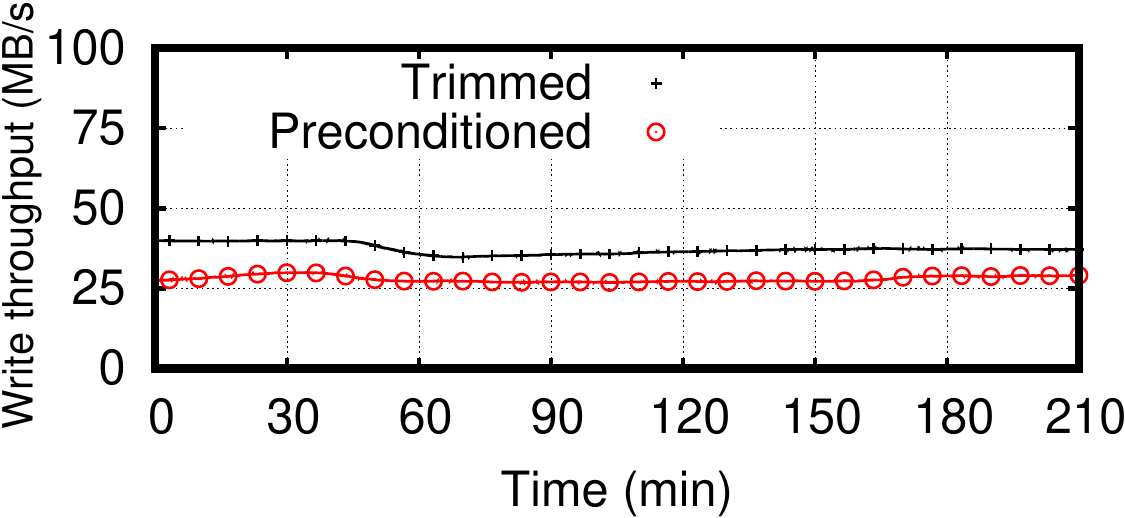}
     }
     }
     \\
       \subfloat[RocksDB: device-level WA.\label{fig:eval:init:rdb:wad}]{
       \includegraphics[scale=0.75]{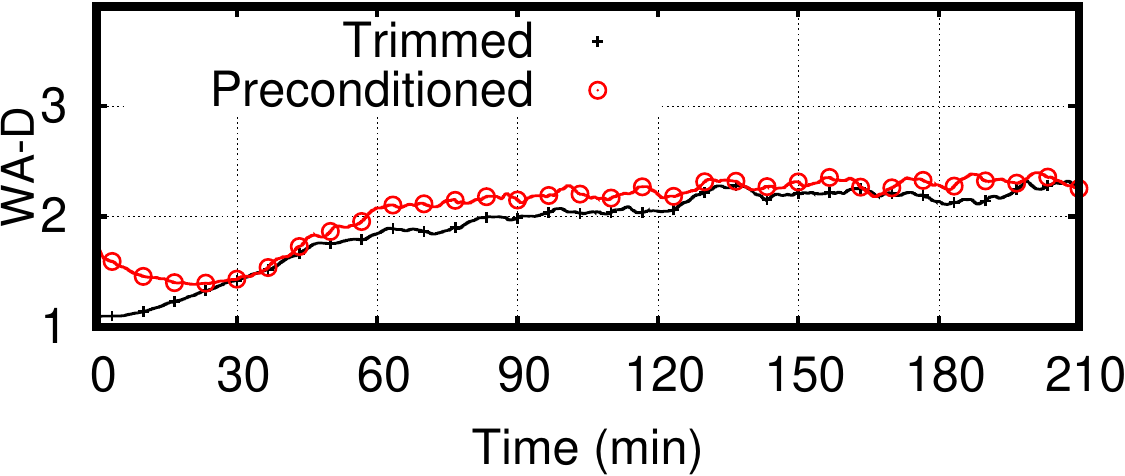}
     } \hfill
     \subfloat[WiredTiger: device-level WA.\label{fig:eval:init:wt:wad}]{
       \includegraphics[scale=0.75]{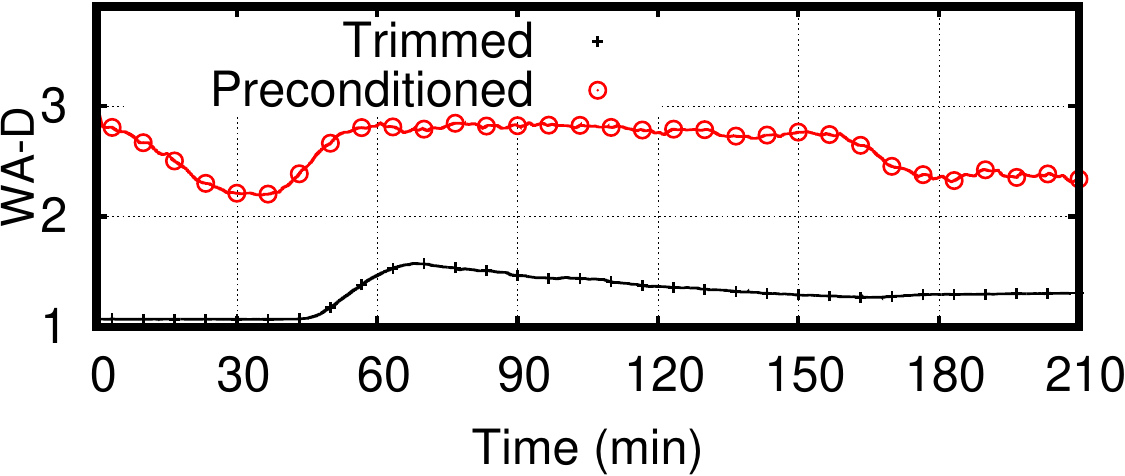}
       }
     \caption{Impact of the initial state of the SSD on performance. Performance achieved over time by RocksDB (left) and WiredTiger (right) depending on the initial conditions of the SSD (trimmed versus preconditioned). The initial conditions of the drive affect throughput (top), potentially even at steady-state, because they affect SSD garbage collection dynamics and the corresponding WA-D (bottom).}
     \label{fig:eval:init}
   \end{figure*}
   
   \begin{figure}[b!]
\centering{
\includegraphics[scale=0.67]{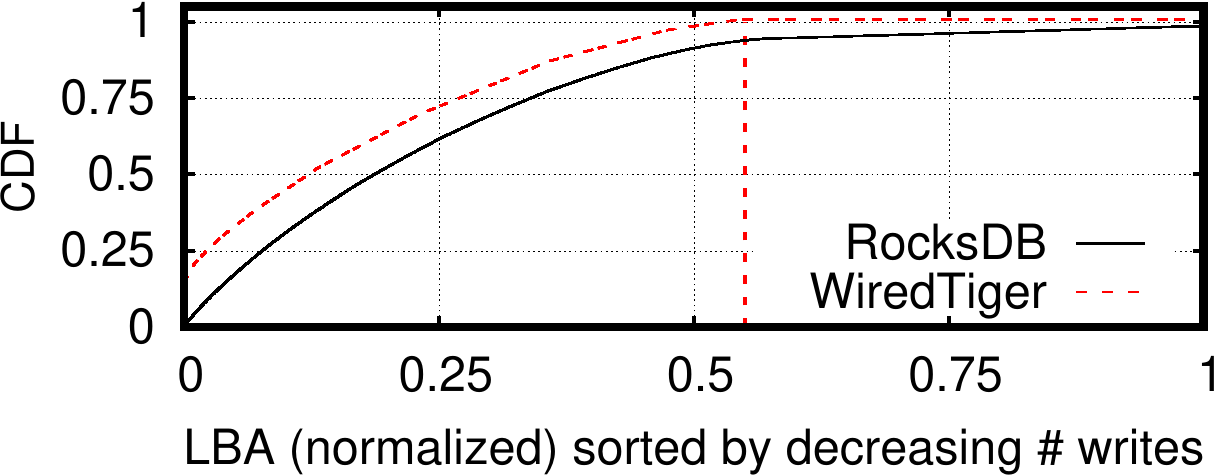}
}
\caption{CDF of the LBA write probability in RocksDB and WiredTiger. The vertical dotted line indicates where the CDF corresponding to WiredTiger reaches value 1, and indicates that WiredTiger does not write to $\approx 45$\% of the LBAs. This write access pattern results into different WA-D depending on the initial state of the drive (see Figure~\ref{fig:eval:init:wt:wad}.)}\label{fig:eval:lba}
\end{figure}
\subsection{Analysis of WA-D} 
\pitfallN{2}{Not analyzing WA-D}{Overlooking WA-D leads to partial or even inaccurate performance analysis.}  %

\evidence{} Many evaluations only consider WA-A in their analysis, which can lead to inaccurate conclusions. We advocate considering WA-D when discussing the performance of a \pts{} for three (in addition to being fundamental to identify steady state, as discussed previously) main reasons.

$i)$ {\em WA-D directly affects the throughput of the device, which strongly correlates with the application-level throughput.}  Analyzing WA-D explains performance variations that cannot be inferred by the analysis of the WA-A alone. Figure~\ref{fig:eval:ss:wt:xput} shows that WiredTiger exhibits a throughput drop at around the 50th minute mark, despite the fact  Figure~\ref{fig:eval:ss:wt:wa} shows no variations in  WA-A. Figure~\ref{fig:eval:ss:wt:wa} shows that at the 50th minute mark WA-D increases from its initial value of 1, indicating that the SSD has run out of clean blocks, and the garbage collection process has started. This increase in WA-D explains the reduction in SSD throughput, which ultimately determines the drop in the throughput achieved by WiredTiger.

The analysis of WA-D also contributes to explaining the performance drops in RocksDB, depicted in Figure~\ref{fig:eval:ss:rdb:xput}. Throughout the test, the KV throughput drops by a factor of $\approx$4,   from 11 to 2.5 KOps/s. Such a huge degradation is not consistent with the   $\approx$2 increase of WA-A and the slightly increased CPU overhead caused by internal LSM-Tree operations (most of the CPUs are idle throughout the test). The doubling of the WA-D explains the huge device throughput degradation, which contributes to the application-level throughput loss.

$ii)$ {\em WA-D is an essential measure of the I/O efficiency of a \pts{}.}  One needs to multiply WA-A by WA-D to obtain the end-to-end write amplification -- from application to memory cells--  incurred by a \pts{} {\em on flash}. This is the write amplification value that should be used to quantify the I/O efficiency of a \pts{} on flash, and its implications on the lifetime of an  SSD. 
Focusing on \wau{} alone, as done in the vast majority of \pts{} performance studies,  may lead to incorrect conclusions. For example, Figure~\ref{fig:eval:ss} (bottom) shows that RocksDB incurs a steady-state \wau{} of 12, which is higher than the \wau{} achieved by WiredTiger by a factor of 1.2$\times$. However, the end-to-end write amplification of RocksDB is 25, which is 2.1$\times$ higher than WiredTiger's.

\begin{figure*}
       \includegraphics[scale=0.65]{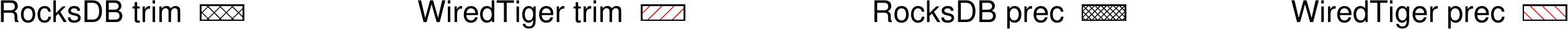}\\
   \subfloat[Throughput.\label{fig:eval:ds:xput}]{
       \includegraphics[scale=0.65]{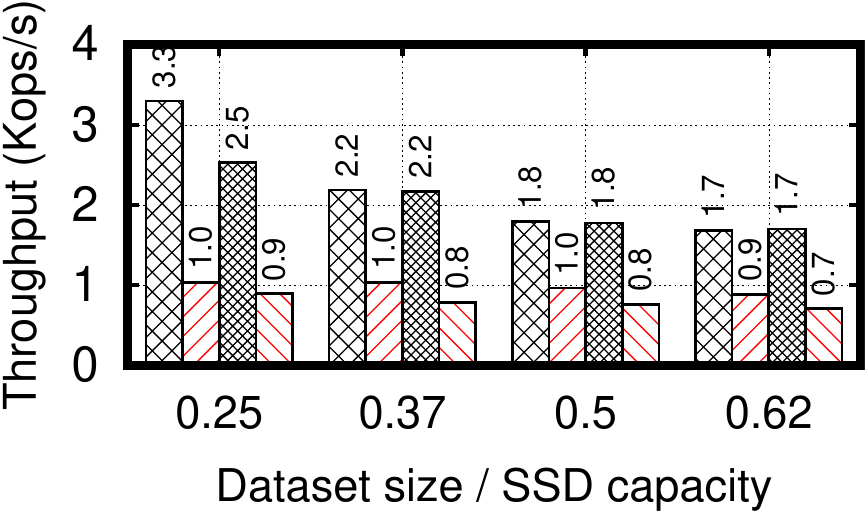}
     } 
     \subfloat[WA-D.\label{fig:eval:ds:wad}]{
       \includegraphics[scale=0.65]{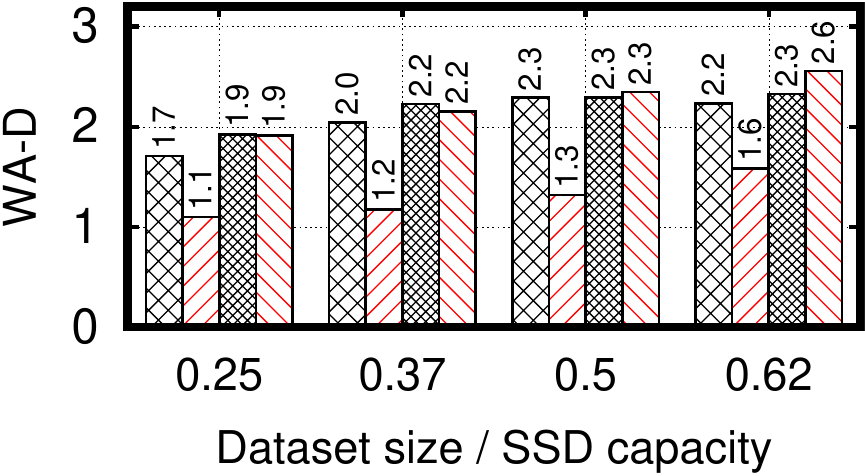}
       }       
     \subfloat[WA-A.\label{fig:eval:ds:wau}]{
       \includegraphics[scale=0.65]{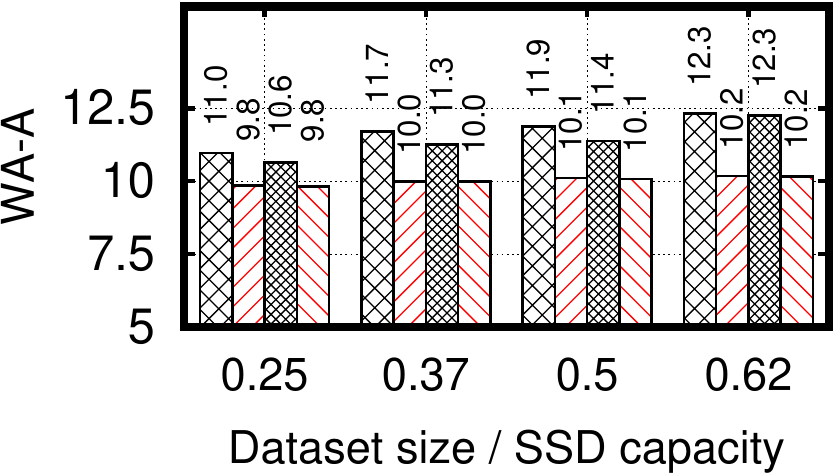}
       }     
     \caption{Impact of the size of the dataset in RocksDB and WiredTiger, with preconditioned and trimmed device. Larger datasets lead to a lower throughput (a). This is mostly due to an increase in WA-D (b), since the WA-A only increases mildly (c).}\label{fig:eval:ds}
   \end{figure*}

$iii)$ {\em WA-D measures the flash-friendliness of a \pts{}.} A low WA-D indicates that a \pts{} generates a write access pattern towards the SSD that does not incur much garbage collection overhead. Measuring WA-D, hence, allows for quantifying the fitness of a \pts{}  for flash SSD deployments, and for verifying the effectiveness of  flash-conscious design choices.

For example, LSM-Trees are often regarded as flash-friendly due to their highly sequential writes, while B+Tree are considered less flash-friendly due to their random write access pattern. The direct measurement of \wad{} in our tests, however, capsizes this conventional wisdom, showing  that RocksDB and WiredTiger achieve a \wad{} of around 2.1 and 1.5, respectively. As a reference, a pure random write workload, targeting also 60\% of the device capacity, has a WA-D of 1.4~\cite{stoica:vldb2013}. In the next section we provide additional insights on the causes of such mismatch between expectation and measured performance.

\guideline{} The analysis of the WA-D  should be a standard step in the performance study of any \pts{}. Such analysis is fundamental to measure properly the flash-friendliness and I/O efficiency of alternative systems. Moreover, the analysis of WA-D leads to important insights on the internal dynamics and performance of a \pts{}, as we show in the following sections.

\subsection{Initial conditions of the drive}\label{sec:eval:ssd}
\pitfallN{3}{Overlooking the internal state of the SSD}{Not controlling the initial condition of the SSD may lead to biased and non-reproducible performance results.} 

\evidence{} Figure~\ref{fig:eval:init} shows the performance over time of RocksDB (left) and WiredTiger (right), when running on an SSD that has been trimmed or preconditioned before starting the experiment. The top row reports KV throughput, and the bottom one reports WA-D.  The plots {\em do not} show the performance of the systems during the initial loading phase.

The plots show that the initial state of the SSD heavily affects the performance delivered by a \pts{} and that, crucially, the steady-state performance of a \pts{} can greatly differ depending on the initial state of the drive.  Such a result is surprising, given that one would expect the  internal state of an SSD to converge to the same configuration, if running the same workload for long enough, and hence to deliver the same steady-state performance regardless of the initial state of the SSD.

To understand the cause of this phenomenon, we have monitored the host write access pattern generated by RocksDB and WiredTiger with \texttt{blktrace}. Figure~\ref{fig:eval:lba} reports the CDF of the access probability of the  page access frequency in the two systems. We observe that in  WiredTiger 46\% of the pages are not written (0 read or write accesses). This indicates that WiredTiger only writes to a limited portion of the logical block address (LBA) space, corresponding to the LBAs that initially store the $\approx 200$ GB of KV pairs plus some small extra capacity, i.e., $\approx 50\%$ of the SSD's capacity in total.
In the case of the trimmed device, this data access pattern corresponds to having only $50\%$ of the LBAs with associated valid data. Because SSD garbage collection only relocates valid data, this gives $\approx 200$ GBs of extra over-provisioning to the  SSD, which in turn leads to a low \wad{}. In a preconditioned device, instead, all LBAs have associated valid data, which means that the garbage collection process has only the hardware over-provisioning available, and needs to relocate more valid pages when erasing a block, leading to a higher \wad{}.

The difference in WA-D, and hence in performance, depending on the initial state of the SSD is much less visible in RocksDB. This is due to the facts that $i)$ the LSM tree utilizes more capacity than a B+Tree and $ii)$ RocksDB writes to the whole range of the LBA space. Hence, the initial WA-D for RocksDB depends heavily on the initial device state, however, all LBAs are eventually over-written and thus the WA-D converges to roughly the same value, regardless of the initial state of the drive.

Our results and analysis lead to two important lessons.

$i)$ The I/O efficiency of a \pts{} on SSD is not only a matter of the {\em high level design} of the \pts{}, but also on its {\em low-level implementation}. Our experiments show that the benefits on WA-D given by the large sequential writes of the LSM implementation of RocksDB are lower than the benefits achieved by the B+Tree implementation of WiredTiger,  despite the fact that WiredTiger generates a more random write access pattern. %

$ii)$ Not controlling the initial state of the SSD can potentially jeopardize two key properties of a \pts{} performance evaluation: fairness and reproducibility. The fairness of a benchmarking process can be compromised by simply running the same workload on two different \ptses{} back to back. The performance of the second \pts{} is going to be heavily influenced by the state of the SSD that is determined by the previous test with the other \pts{}. The lack of fairness can lead a performance engineer to pick a sub-optimal \pts{} for their workload, or a researcher to report incorrect results.

The reproducibility of a benchmarking process can be compromised because running two independent tests of  a \pts{} with the same workload and on the same hardware may lead to substantially different results.  For production engineers this means that the performance study taken on a test machine may differ widely with respect to the performance observed in production. For researches, it means that it may be impossible to replicate the results published in another work.

\begin{figure*}
\includegraphics[scale=0.65]{gnuplot/size_labels}\\
  \subfloat[Space utilization.\label{fig:eval:sa:df}]{
       \includegraphics[scale=0.65]{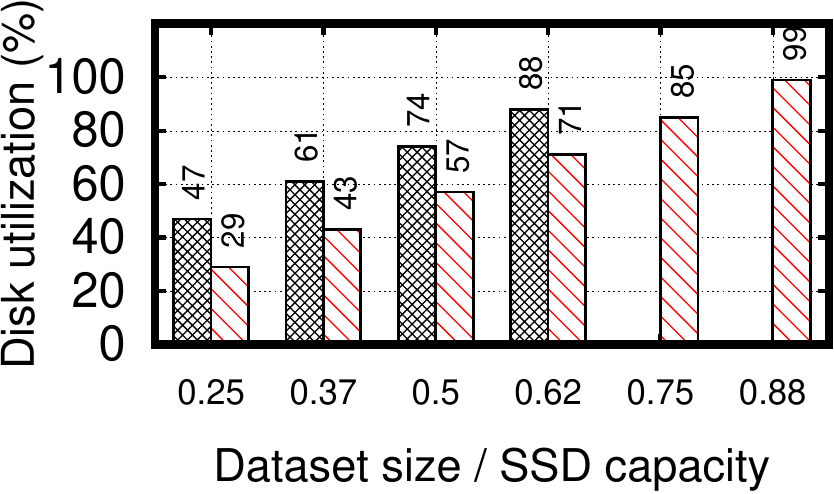}
       }    
  \subfloat[Space amplification.\label{fig:eval:sa:sa}]{
       \includegraphics[scale=0.65]{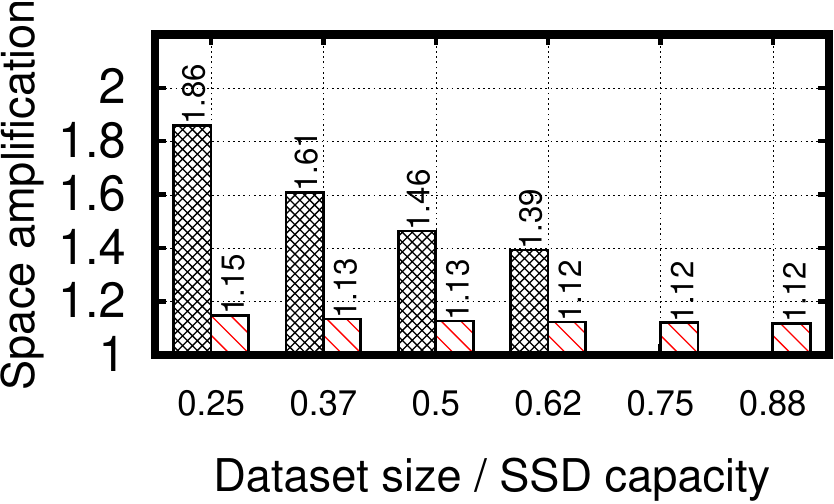}
       }    
  \subfloat[Storage cost comparison.\label{fig:eval:sa:cost}]{
       \includegraphics[scale=0.65]{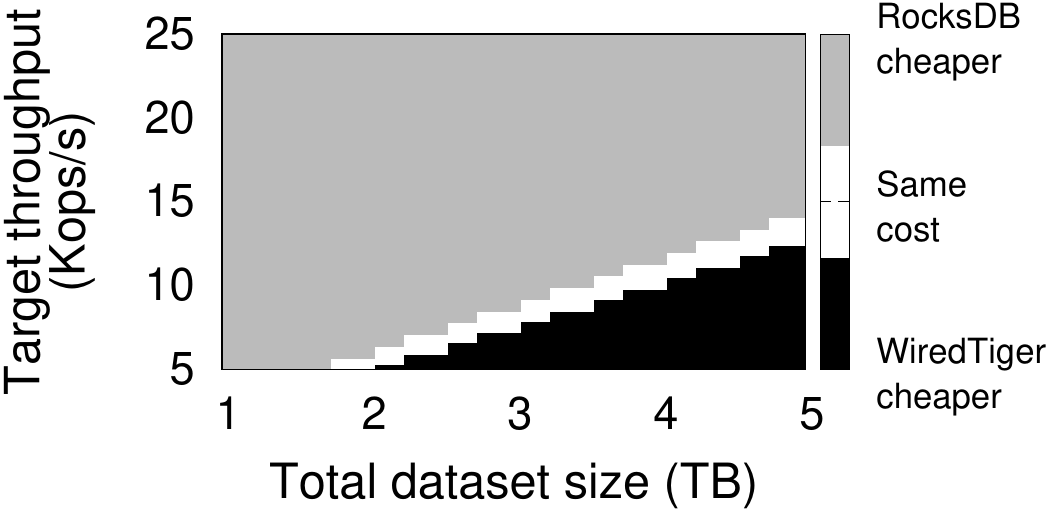}
       }\caption{Space amplification in RocksDB and WiredTiger (trimmed and preconditioned SSD), and its effects on storage costs (preconditioned SSD). RocksDB runs out of space in the two largest datasets we consider. RocksDB uses more space than WiredTiger to store a dataset of a given size (a), leading to a higher space amplification (b). The heatmap (c) reports the system that requires fewer drives  to store a given dataset  while achieving a target throughput --hence incurring a lower storage monetary cost.}
       \end{figure*}

\guideline{} To overcome the aforementioned issues, we recommend to control and report the initial state of the SSD before {\em every} test.  This state depends on the target deployment of the \pts{}. For a performance engineer, such state should be  as similar as possible to the one observed in the production environment, which also depends on other applications possibly collocated with the \pts{}. We suggest to researchers to precondition the SSD as described in Section~\ref{sec:testbed:state}. In this way, they can evaluate the \pts{} in the most general case possible, thus broadening the scope of their results. To save on the preconditioning time, the SSD can be trimmed, provided that one checks that the steady-state performance of the \pts{} does not substantially differ from the one observed on a preconditioned drive.

\subsection{Data-set size}
\pitfallN{4}{Testing with a single dataset size}{The amount of data stored by the SSD changes its behavior and affects overall performance results.}

\evidence{}  Figure~\ref{fig:eval:ds} reports the steady-state throughput (left), WA-D (middle), and \wau{} (right) of RocksDB and WiredTiger with datasets whose sizes span from 0.25 to 0.88 of the capacity of the SSD (from 100GB to 350GB  of KV pairs). We do not report results for RocksDB for the two biggest datasets because it runs out of space. The figure reports results both with a trimmed and with a preconditioned SSD. 

Figure~\ref{fig:eval:ds:xput} shows that the throughput achieved by the two systems is affected by the size of the dataset that they manage, although to a different extent and in different ways depending on the initial state of the SSD. By contrasting Figure~\ref{fig:eval:ds:wad}  and Figure~\ref{fig:eval:ds:wau} we conclude that the performance degradation brought by the larger data-set is primarily due to the  idiosyncrasies of the SSD. In fact, larger datasets lead to more valid pages in each flash block, which increases the amount of data being relocated upon performing garbage collection, i.e., the WA-D.

Changing the dataset size affects the comparison between the two systems both quantitatively and qualitatively. We also note that the comparison among the two systems is affected by the initial condition of the drive.

 On a trimmed SSD, RocksDB achieves a throughput that is 3.3$\times$ higher than WiredTiger's when evaluating the two systems on the smallest dataset. On the largest dataset, however, this performance improvement shrinks to 1.9$\times$.
Moreover, WiredTiger exhibits a lower WA-D across the board, due to the LBA access pattern discussed in the previous section.

On a preconditioned SSD, the speedup of RocksDB over WiredTiger still depends on the size of the dataset, but it is lower in absolute values than on a trimmed SSD, ranging from 2.7$\times$ on the smallest dataset to 2.57$\times$ on the largest one. Moreover, whether RocksDB has a better WA-D than  WiredTiger depends on the  dataset size. In particular, the WA-D of RocksDB and WiredTiger are approximately equal when storing datasets whose sizes are up to half of the drive's capacity. Past that point, RocksDB's WA-D is sensibly lower than WiredTiger's (2.3 versus 2.6). This happens because the benefits of WiredTiger's LBA access pattern decrease with the size of the dataset (and hence of the range of LBAs storing KV data) and the reduced over-provisioning due to preconditioning.

\guideline{}  We suggest production engineers to benchmark alternative \ptses{} with a dataset of the size that is expected in production, and refrain from testing with scaled-down datasets for the sake of time. 
We suggest researchers to experiment with different dataset sizes. This suggestion has a twofold goal. First, it allows a researcher to study the sensitivity of their \pts{} design to different device utilization values. Second, it disallows evaluations that are purposely or accidentally biased in favor of one design over another.

\subsection{Space amplification}
\pitfallN{5}{Not accounting for space amplification}{The space utilization overhead of a \pts{} determines its storage requirements and deployment monetary cost.}

\evidence{} \ptses{} frequently trade additional space for improved performance, and understanding their behavior depends on understanding these trade-offs. Figure~\ref{fig:eval:sa:df} reports the total disk utilization incurred by RocksDB and WiredTiger depending on the size of the dataset. The disk utilization includes the overhead due to filesystem meta-data. Because RocksDB frequently writes and erases many large files, its disk utilization varies sensibly over time. The value we report is the maximum utilization that RocksDB achieves. Figure~\ref{fig:eval:sa:sa} reports the space amplification corresponding to the utilization depicted in Figure~\ref{fig:eval:sa:df}. 

WiredTiger uses an amount of space only slightly higher than the bare space needed to store the dataset, and achieves a space amplification that ranges from 1.15 to 1.12. RocksDB, instead, requires much more additional disk space to store the several levels of its LSM-Tree. Overall, RocksDB achieves an application space amplification ranging between 1.86, with the smallest dataset we consider, to 1.39, with the biggest dataset that it can handle~\footnote{\scriptsize The disk utilization in RocksDB depends on the setting of its internal parameters, most importantly, the maximum number of levels, and the size of each level~\cite{rocksdb-tuning}. It is possible to achieve a lower space amplification  than the one we report, but at the cost of substantially lower throughput due to increased compaction overhead.}.

These results show that space amplification plays a key role in the performance versus storage space trade-off. Such trade-off affects
the total storage cost of a \pts{} deployment, given an SSD drive model, a target throughput, and total dataset size. In fact, a \pts{} with a low space amplification may fit the target dataset in a smaller and cheaper drive with respect to another \pts{} with a higher write amplification, or can index more data given the same capacity, requiring fewer drives to store the whole dataset. 

To showcase this last point, we perform a back-of-the-envelope computation to identify which of the two systems require fewer SSDs (and hence incur a lower storage cost) to store a given dataset and at the same time achieve a target throughput. We use the throughput and disk utilization values that we measure for our SSD (see Figure~\ref{fig:eval:ds:xput} and Figure~\ref{fig:eval:sa:df}). For simplicity, we assume one \pts{} instance per SSD, and  that the aggregate throughput of the deployment is the sum of the throughputs of the instances. 
Figure~\ref{fig:eval:sa:cost} reports the result of this computation. Despite having a lower per-instance throughput, the higher space efficiency of WiredTiger makes it preferable over RocksDB in scenarios with a large dataset and a relatively low target throughput. This configuration represents an important class of workloads, given that with the ever-increasing amount of data being collected and stored, many applications begin to be storage capacity-bound rather than throughput-bound~\cite{cidon:atc:2017}.

\guideline{} The experimental evaluation of a \pts{} should not focus only on performance, but should include also space amplification. For research works, analyzing space amplification provides additional insights on the performance dynamics and trade-offs of the design choices of a \pts{},  and allows for a multi-dimensional comparison among designs. For production engineers, analyzing space amplification is key to  compute the monetary cost of provisioning the storage for a \pts{} in production, which is typically more important than sheer maximum performance~\cite{microsoft:ssd}.

\begin{figure}[t!]
       \includegraphics[scale=0.65]{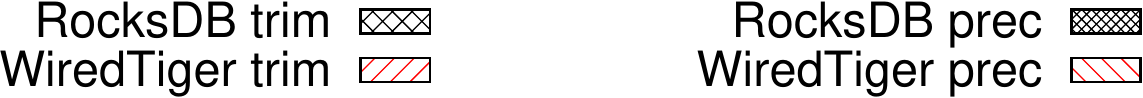}
   \subfloat[Throughput.\label{fig:eval:op:xput}]{
       \includegraphics[scale=0.65]{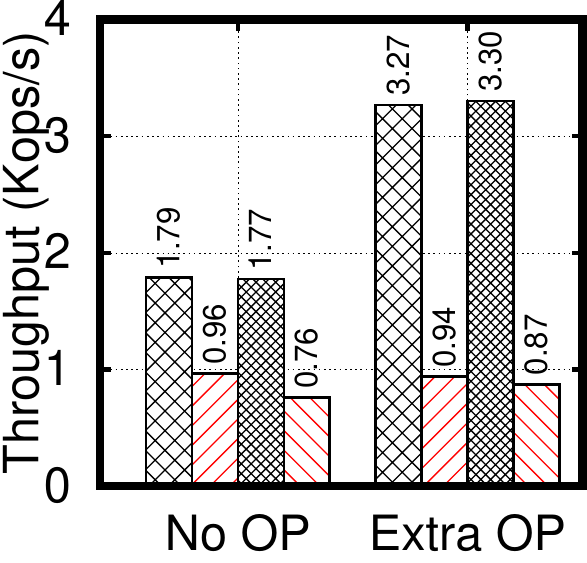}
     } 
     \subfloat[WA-D.\label{fig:eval:op:wad}]{
       \includegraphics[scale=0.65]{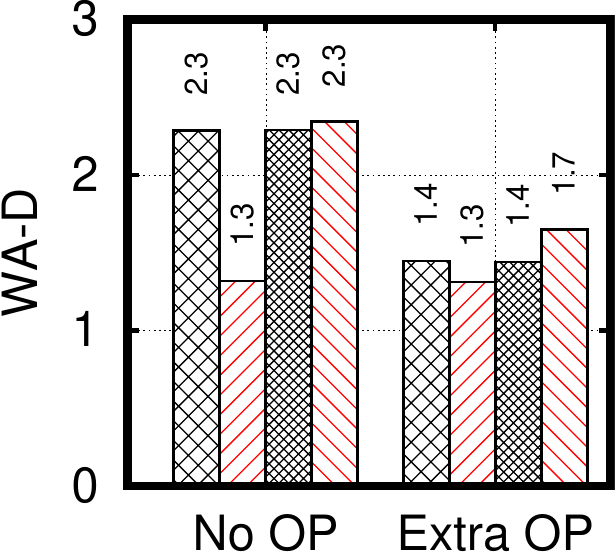}
       }      
     \caption{Impact of using extra SSD over-provisioning (OP). Extra OP may increase throughput (a) by improving WA-D (b), at the cost of reducing the amount of data that the SSD can store.} \label{fig:eval:op}
   \end{figure}

As a final remark, we note that this pitfall applies also to \ptses{} not deployed over an SSD, and hence our considerations apply more broadly to \ptses{} deployed over any persistent storage medium.

\subsection{SSD over-provisioning}
\pitfallN{6}{Overlooking SSD software over-provisioning}{Allocating extra over-provisioning capacity to the SSD may lead to more favorable capacity versus performance trade-offs.}

\evidence{} Figure~\ref{fig:eval:op} compares the steady-state throughput (left) and WA-D (right) achieved by RocksDB and WiredTiger in two settings: $i)$ the default one in which the whole SSD capacity is assigned to the disk partition accessible by the filesystem underlying the \pts{}, and $ii)$ one in which some SSD space is not made available to the filesystem underlying the \pts{}, and is instead assigned as extra over-provisioning to the  SSD. Specifically, in the second setting we trim the SSD and assign a 300GB partition to the \pts{}. Hence, the SSD has 100GB of trimmed capacity that is not visible to the \pts{}. We choose this value because 100GB corresponds to half of the free capacity of the drive once the 200 GB dataset has been loaded. For both settings we consider the case in which the \pts{} partition remains clean after the initial trimming, and the case in which it is preconditioned.

Extra over-provisioning  improves the performance of RocksDB by a factor of 1.83$\times$. This substantial improvement is caused by a drastic reduction of WA-D, that drops from 2.3 to 1.4, and it applies to RocksDB regardless of the initial state of the \pts{} partition, for the reason discussed in Section~\ref{sec:eval:ssd}. 

\begin{figure}[t!]
   \centering
   \includegraphics[scale=0.8]{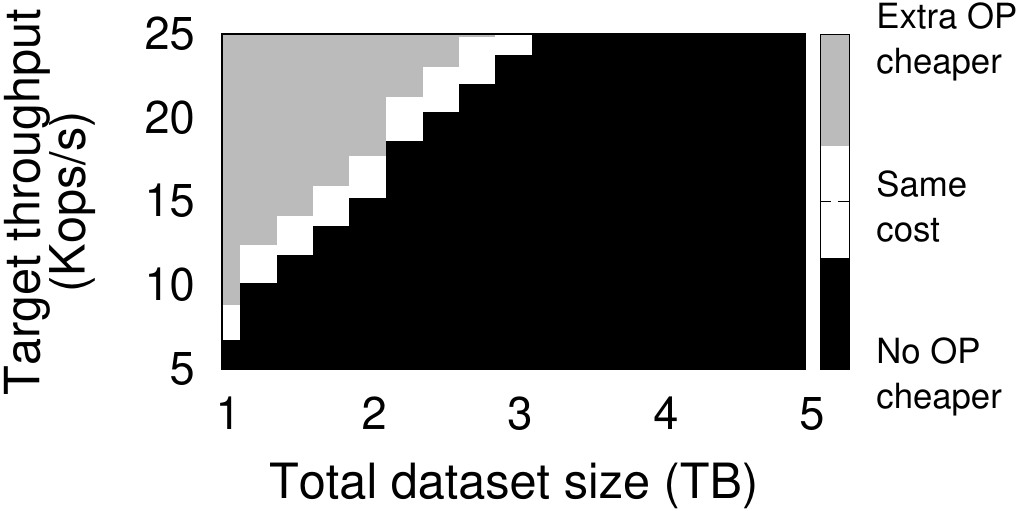}\caption{Storage cost comparison of using RocksDB with or without extra over-provisioning (OP) on a preconditioned SSD. The heatmap reports the RocksDB configuration that requires fewer drives  to store a given dataset  while achieving a target throughput --hence incurring a lower storage monetary cost.}\label{fig:eval:op:rdb}
   \end{figure}

\begin{figure}[b!]
\includegraphics[scale=0.7]{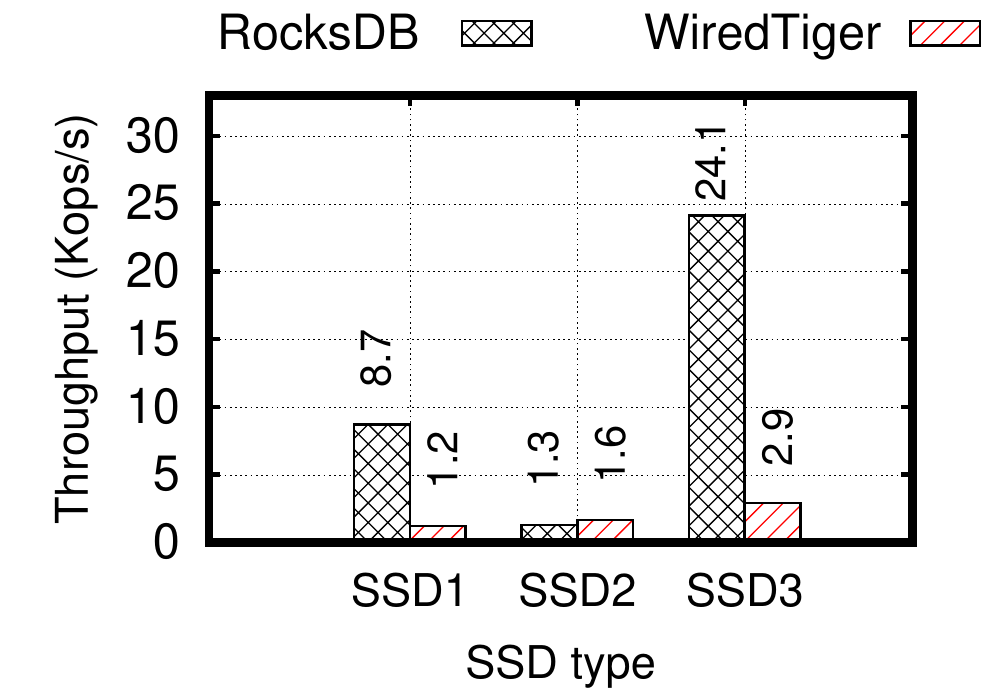}
\caption{Impact of the SSD type on the throughput of RocksDB and WiredTiger. The type of SSD significantly affects the absolute performance achieved by the two systems, and can even determine which of them achieves the higher throughput.}\label{fig:ssd_impact_on_xput}
\end{figure}

\begin{figure*}[t!]
 \centering{\includegraphics[scale=0.72]{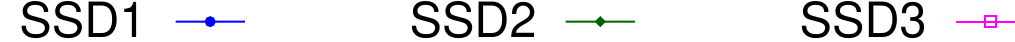}}\\
	\subfloat[RocksDB.\label{fig:ssd_impact_on_qos_rdb}]{
		\includegraphics[scale=0.7]{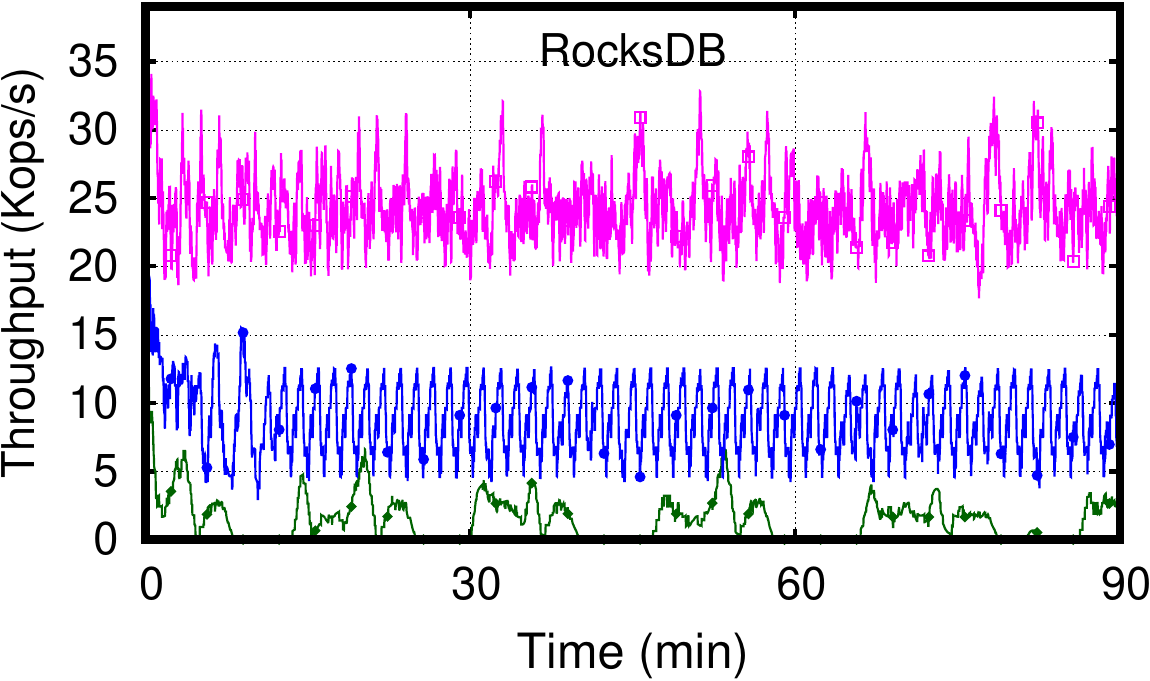}
	}      
	\subfloat[WiredTiger.\label{fig:ssd_impact_on_qos_wt}]{
		\includegraphics[scale=0.7]{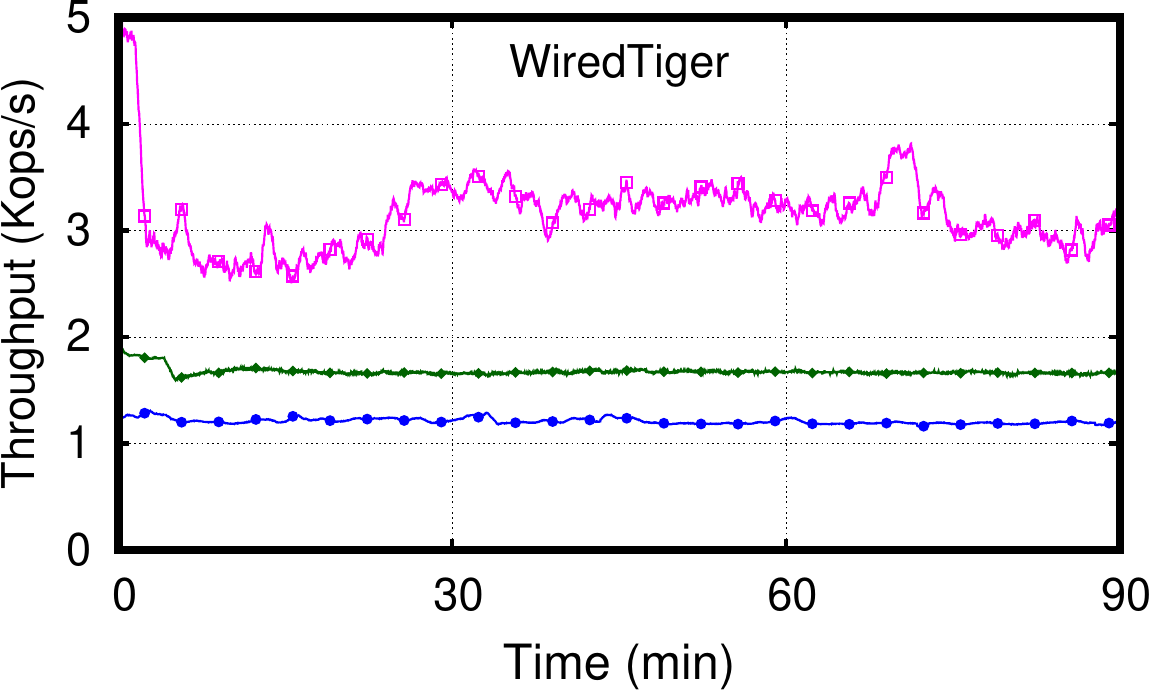}
	}     
	\caption{Throughput (1 minute average) of RocksDB (left) and WiredTiger (right) over time when running on SSDs based on different technologies. The type of SSD influences heavily the throughput variability of RocksDB, and much less that of WiredTiger.} \label{fig:ssd_impact_on_qos}
\end{figure*}

The impact of extra over-provisioning  is much less evident in WiredTiger. In the trimmed device case, the extra over-provisioning has no effect on  WiredTiger. This happens because WiredTiger writes only to a certain range of the LBA space (see Figure~\ref{fig:eval:lba}). Hence, all other trimmed blocks act as over-provisioning, regardless of whether they belong to the \pts{} partition or the extra over-provisioning one. On the preconditioned device, instead, all blocks of the \pts{} partition  have data associated with them, so the only software over-provisioning  is given by the trimmed partition. This extra over-provisioning reduces WA-D from 1.7 to 1.3, yielding a throughput improvement of 1.14$\times$.

Allocating extra over-provisioning can be an effective technique to reduce the number of \pts{} instances needed in a deployment (and hence reduce its storage cost), because it increases the throughput of the \pts{} without requiring additional hardware resources. However, extra over-provisioning also reduces the amount of data that a single drive can store, which potentially increases the amount of drives needed to store a dataset and the storage deployment cost.  
 To assess in which use cases using extra over-provisioning is the most cost-effective choice we  perform a back-of-the-envelope computation of the number of drives needed to provision a RocksDB deployment  given a dataset size and a target throughput value.  We perform this study on RocksDB because it benefits the most from extra over-provisioning. We use the same simplifying assumptions that we made for the previous similar analysis. Figure~\ref{fig:eval:op:rdb} reports the results of our computation. As expected, extra over-provisioning is beneficial for use cases that require high throughput for relatively small datasets. For larger datasets with relatively low performance requirements, it is more convenient to allocate as much of the drive's capacity as possible to RocksDB.

\guideline{} It is well known that \ptses{} have several tuning knobs that have a huge effect on performance~\cite{lim:fast:2016,lsm-bush:sigmod19,vat:2020}. We suggest to consider SSD over-provisioning as an additional, yet first class tuning knob of a \pts{}.  SSD extra over-provisioning trades capacity for performance, and can reduce the storage cost of a \pts{} deployment in some use cases.

\subsection{Storage technology}
\pitfallN{7}{Testing on a single SSD type}{The performance of a PTS heavily depends on the type of SSD used. This makes it hard to extrapolate the expected performance when running on other drives and also to reach conclusive results when comparing two systems.}

\evidence{}
We exemplify this pitfall through an experiment where we keep the workload and the \RDB and \WT configurations constant and only swap the underlying  storage device. 
We use three SSDs:  an Intel p3600~\cite{intel:p3600} flash SSD, i.e., the drive used for the experiments discussed in previous sections; an Intel 660~\cite{intel-660p} flash SSD; and an Intel Optane~\cite{optane}. We refer to these SSDs as SSD1, SSD2 and SSD3, respectively, in the following discussion. SSD3 is a high-end SSD, based on the newer 3DXP technology that achieves higher performance than flash SSDs. We use SSD3 as an upper bound on performance that a \pts{} can achieve on a flash block device.

To try and isolate the performance impact due to the SSD technology (i.e., architecture and underlying storage medium performance) itself in the assessment of a \pts{}, we eliminate, as much as possible, the other sources of performance variability that we have discussed so far. To this end, we run a workload with a dataset that is 10$\times$ smaller than the default one, and we trim the flash SSDs. In this way, the effect of garbage collection is the flash SSDs is minimized, resulting in a WA-D very close to one. %

\begin{figure*}
  {\begin{center}
    \includegraphics[scale=0.75]{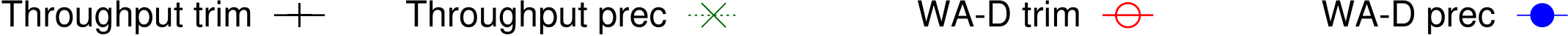}\\
    \end{center}}
    
    \begin{center}
   \subfloat[RocksDB, 50:50 r:w ratio.\label{fig:eval:5050:rdb}]{
       \includegraphics[scale=0.75]{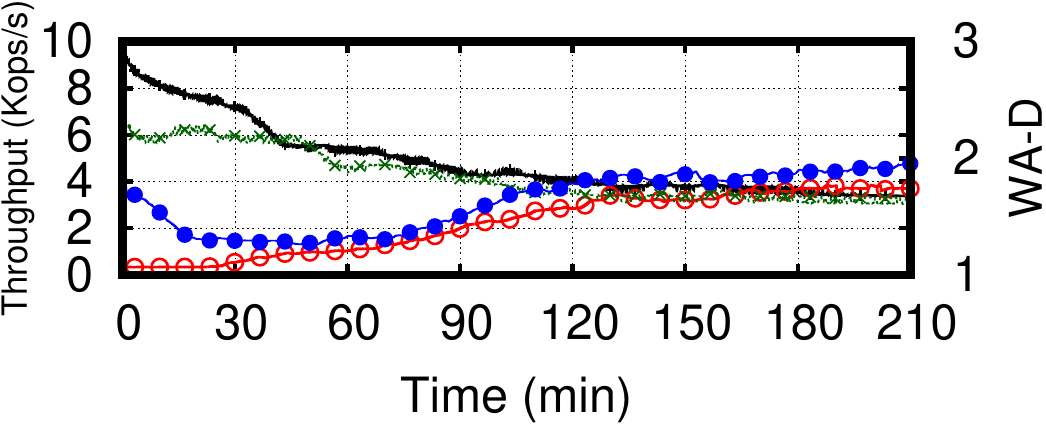}
     }      \hfill
     \subfloat[WiredTiger, 50:50 r:w ratio.\label{fig:eval:5050:wt}]{
       \includegraphics[scale=0.75]{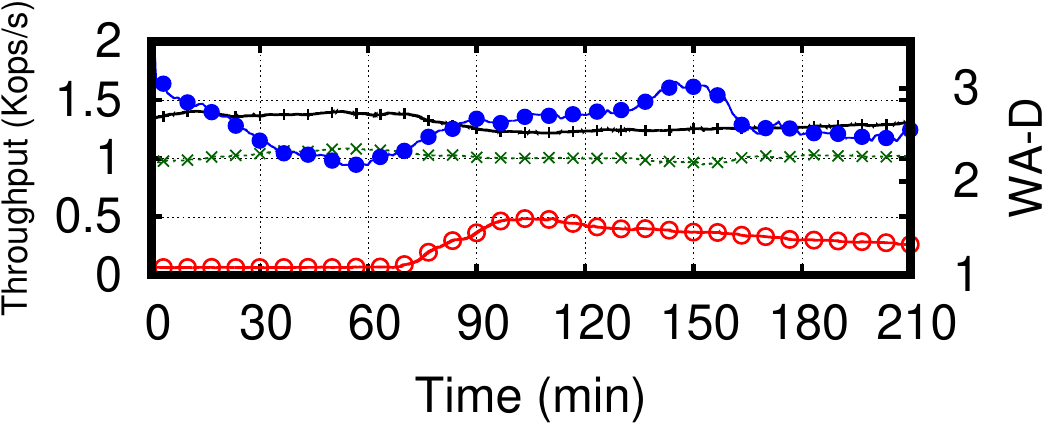}
       }     \\
       \end{center}
       
       \begin{center}
        \subfloat[RocksDB, 128B values.\label{fig:eval:small:rdb}]{
       \includegraphics[scale=0.75]{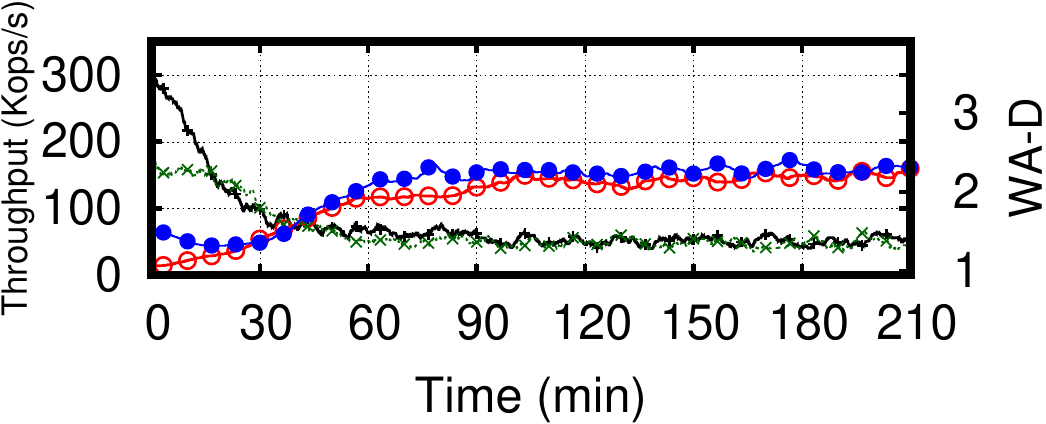}
     } \hfill
     \subfloat[WiredTiger, 128B values. \label{fig:eval:small:wt}]{
       \includegraphics[scale=0.75]{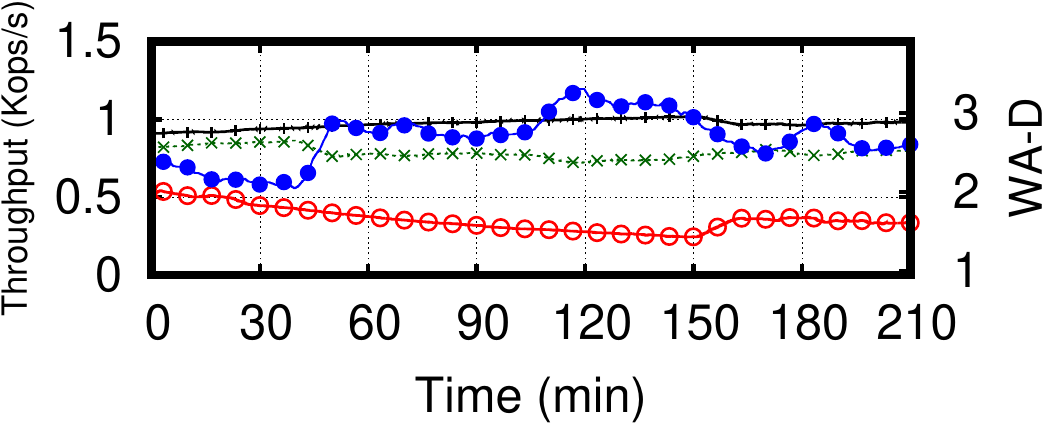}       
       }
       \end{center}
           \caption{Performance of RocksDB (left) and WiredTiger (right) over time, with a preconditioned and trimmed device. The top row refers to a workload with small (128 bytes) values. The bottom row refers to a workload with a 50:50 read:write ratio. The pitfall we describe  apply to a broad set of workloads as long as they have a sustained write component (here we represent pitfalls \#1, \#2 and \#3).} \label{fig:eval:other} 
   \end{figure*}

Figure~\ref{fig:ssd_impact_on_xput} shows the steady-state throughput of RocksDB and WiredTiger when deployed on the three SSDs. As can be depicted from the plot, the performance impact changing the underlying drive varies drastically across the two systems. Explaining these performance variations requires gaining a deeper understanding of the low-level design of the SSDs, which is not always achievable. %

RocksDB achieves the highest throughput on SSD3, and a higher throughput on SSD1 than on SSD2. This performance trend is mostly determined by the write latencies of the SSDs, which are the lowest in SSD3, and lower in SSD2 than in SSD1.
Also WiredTiger achieves the highest throughput on SSD3 but, surprisingly, it obtains a higher throughput on SSD2 than on SSD1. We argue that the reason for this performance dynamic lies in the fact that SSD2 has a larger internal cache than SSD1. Because \WT performs small writes, uniformly distributed over time, the cache of SSD2 is able to absorb them with very low latency, and destages them in the background. The larger cache of SSD2 does not yield the same beneficial effects to \RDB because \RDB performs large  bursty writes, which overwhelm the cache, leading to longer write latencies and, hence, lower throughput.

These dynamics also lead to the surprising result that either of the two systems we consider can achieve a higher throughput than the other, just by changing the SSD on which they are deployed.  

We also observe very different performance variations for the two systems, when deployed on different SSDs.  On the one hand, the best and worst throughputs achieved by RocksDB vary by a factor of almost 20$\times$ (SSD2 versus SSD3). On the other hand, they vary only by a factor of 2.4 for WiredTiger.

These results are especially important, because they indicate that the performance comparison across \pts{} design points, and the corresponding conclusions that are drawn, are strongly dependent on the SSD employed in the benchmarks, and hence hard to generalize.

The type of SSD also affects dramatically the performance predictability of the two system.  Figure \ref{fig:ssd_impact_on_qos} reports the throughput of RocksDB (left) and WiredTiger (right) when deployed over the three SSDs. To highlight the performance variability we average the throughput over a 1 minute interval (as opposed to the default 10 minutes used in previous plots). 

The throughput of \RDB  varies widely over time, and the extent of the variability depends on the type of SSD.
When using SSD1, \RDB suffers from throughput swings of $100\%$. When using SSD2, \RDB has long periods of time where no application-level writes are executed at all. This happens because the large writes performed by \RDB overwhelm the cache of SSD2 and lead to long stall times due to internal data destaging. On SSD3, the relative \RDB throughput variability decreases to $30\%$.
\WT is less prone to performance variability, and exhibits a more steady and predictable performance irrespective of the storage technology.

\guideline{}  We have shown that it is difficult to predict the performance of a \pts{}  just based on the high-level SSD specifications (e.g., bandwidth and latency specifications). Therefore, we recommend  testing a \pts{} on multiple SSD classes, preferably using devices from multiple vendors, and using multiple flash storage technologies. 
Testing on multiple SSD classes allows researchers to 
draw broader and more significant conclusions about the performance of a \pts{}'s design, and to assess the ``intrinsic" validity of such design, without tying it to specific characteristics of the medium on which the design has been tested.
For a performance engineer, testing with multiple drives is essential to identify which one yields the best combination of storage capacity, performance and cost depending on the target workload.

\subsection{Additional workloads}
\label{sec:eval:other}
In this section, we show that our pitfalls affect a wider set of workloads than the one we have considered so far.
For space constraints, we focus on the first three pitfalls.
Figure~\ref{fig:eval:other} reports the throughput and WA-D over time achieved by RocksDB (left) and WiredTiger (right) with two workloads that we obtain by varying one parameter of our default workload. The plots {\em do not} show the performance of the systems during the initial loading phase.
The top row refers to a workload in which the size of the values is 128 bytes (vs. 4000 used so far). To keep the amount of data stored indexed by the \pts{} constant to the one used by the previous experiments, we increase  accordingly the number of keys. The bottom row refers to a workload in which the top level application submits a mix of read and write operations, with a 50:50 ratio. We run these workloads using both a preconditioned SSD and a trimmed one.

As the plots show, the first three pitfalls apply to these workloads as well.  First, steady-state behavior can be very different from the performance observed at the beginning of the test. In the mixed read/write workload, it takes longer to the \ptses{} and to the SSD to stabilize, because writes are less frequent. This is visible by comparing the top row in Figure~\ref{fig:eval:other} with the top row in Figure~\ref{fig:eval:ss} (Section~\ref{sec:eval:ss}).

Second, the value and the variations of WA-D are important to explain performance changes and dynamics. We note that the WA-D function of WiredTiger in the trimmed case with small values (Figure~\ref{fig:eval:small:wt}) is different from the one observed for the workload with 4000B values  (Figure~\ref{fig:eval:ss:wt:wa} in Section~\ref{sec:eval:ss}). With 4000B values the WA-D function starts at a value close to 1, whereas with 128B values its starting point is closer to 2. This happens because the  initial data loading phase leads to different SSD states depending on the size of the KV pairs. With 4000B values, one KV pair can fill one filesystem page, which is written only once to the underlying SSD. With small values, instead, the same page needs to written multiple times to pack more KV pairs, which  in higher fragmentation at the SSD level.   Such a phenomenon is not visible in RocksDB, because it writes large chunks of data regardless the size of the individual KV pairs.

Third, the initial state of the SSD leads to different transient and steady-state performance. As for the other workloads considered in the paper, this pitfall applies especially to WiredTiger.

In light of these results, we argue that our pitfalls and guideline should apply to every workload that has a sustained write component. Some of our pitfalls are also relevant to read-only workloads, i.e., the ones concerning the data-set size, the space amplification, and the storage technology.

%% file: related.tex
\section{Related Work}%
\label{sec:rw}
In this section we discuss $i)$ the performance analyses of existing storage systems on SSD, and how they relate to the pitfalls we describe; $ii)$ work on modeling and benchmarking SSDs in the storage community; and $iii)$ research in the broader field of system benchmarking. 

\mypar{Performance analyses of SSD-based storage systems.} Benchmarking the performance of \ptses{} on SSDs is a task frequently performed both in academia~\cite{balmau:atc:2019,sears:sigmod:2012,balmau:atc:2017,bindschaedler:asplos:2020,raju:sosp17:pebblesdb,luo:vldb:2019,dayan:monkey:2018,dayan:dostoevsky:2018,bortnikov:vldb:2018,marmol:atc:2015,lim:sosp:2011}  and in industry~\cite{leveldb-bench,hyperleveldb-bench,wiredtiger-bench,rocksdb-bench} to compare different designs. 

In general, these evaluations fall short in considering the benchmarking pitfalls we discuss in this paper. For example, the evaluations of the systems do not report the duration of the experiments, or they do not specify the initial conditions of the SSD on which experiments are run, or consider a single dataset size. As we show in this paper, these aspects are crucial for both the quantitative and the qualitative analysis of the performance of a data store deployed on an SSD.

In addition, performance benchmarks of \ptses{} typically focus on application-level write amplification to analyze the I/O efficiency and flash-friendliness of a system~\cite{lim:sosp:2011,balmau:atc:2017,raju:sosp17:pebblesdb,dayan:monkey:2018,dayan:dostoevsky:2018,marmol:atc:2015}. We show  that also device-level write amplification must be taken into account for these purposes.

A few systems distinguish between bursty and sustained performance, by investigating the variations of throughput and latencies over time in LSM-tree key value stores~\cite{luo:vldb:2019,sears:sigmod:2012,balmau:atc:2019}. Balmau et al.~\cite{balmau:atc:2019} additionally show how some optimizations can improve the performance of LSM-Tree key-value stores in the short term, but lead to performance degradation in the long run. These works focus on the high-level, i.e., data-structure specific, causes of such performance dynamics. Our work, instead, investigates the low-level causes of performance variations in \ptses{}, and correlates them with the  idiosyncratic performance dynamics of SSDs. 

Yang et al.~\cite{yang:inflow:2014} show that stacking log-structured data structures on top of each of each other may hinder the effectiveness of the log-structured design.  Oh et al.~\cite{oh:fast:2012} investigate the use of SSD over-provisioning to increase the performance of a  SSD-based cache.  Athanassoulis et al.~\cite{rum} propose the RUM conjecture, which states that \ptses{} have an inherent trade-off  between performance and storage cost. 
Our paper touches some of these issues, and complements the findings of these works, by covering in a systematic fashion  several pitfalls of benchmarking \ptses{} on SSDs, and by providing experimental evidence for each of them.

\mypar{SSD performance modeling and benchmarking.}  The research and industry storage communities have produced much work on modeling and benchmarking the performance of SSDs.  
The Storage Networking Industry Association has defined the Solid State Storage Performance Test Specification~\cite{snia}, which contains the guidelines on how to perform rigorous and reproducible SSD benchmarking, so that performance results are comparable across vendors.  Many analytical models~\cite{Hu:systor:2009,stoica:vldb2013,desnoyer:tocs:2014,stoica:mascots:2019} aim to express in closed form the performance and the device level WA of an SSD as a function of the workload and the SSD internals and parameters. MQSim~\cite{tavakkol:fast:2018} is a simulator specifically designed to replicate quickly and accurately the behavior of an SSD at steady state.

Despite this huge body of work, we have previously shown that practitioners and researchers in the system and database communities do not properly  --or entirely--take into account the performance dynamics of SSDs when benchmarking \ptses{}.  With our work, we aim to raise awareness about the SSD performance intricacies in the system and database communities as well. To this end, we show the intertwined effects of the SSD idiosyncrasies on the performance of  \ptses{}, and provide guidelines on how to conduct a more rigorous and SSD-aware performance benchmarking.

\mypar{System benchmarking.} The process of benchmarking a system is notoriously difficult, and can incur subtle pitfalls that may undermine its results and conclusions.  Such a complexity is epitomized by the list of {\em benchmarking crimes}~\cite{benchmarking-crimes}, a popular collection of benchmarking errors and anti-patterns that are frequently found in the evaluation of research papers.  Raasveldt et al.~\cite{raasveldt:2018} provide a similar list with a focus on DB systems. Many research papers target different aspects of the process of benchmarking a system. Mariq et al.~\cite{mariq:osdi:2018}, Uta et al.~\cite{uta:nsdi:2020}, and Hoefler and Belli~\cite{hoefler:sc:2015} focus on the statistical relevance of the measurements, investigating whether experiments can be repeatable, and how many trials are needed to consider a set of measurements meaningful. Our work is complementary to this body of research, in that  it aims to provide guidelines to obtain a more fair and reproducible performance assessment of \ptses{} deployed on SSDs. 

%% file: conclusion.tex
\section{Conclusions}
\label{sec:conclusion}
The complex interaction between a persistent tree data structure and a flash SSD device can easily lead to inaccurate performance measurements. 
In this paper we show seven pitfalls that one can incur when benchmarking a persistent tree data structure on a flash SSD. We demonstrate these pitfalls using RocksDB and WiredTiger, two of the most widespread implementations of the LSM-tree and of the B+tree persistent data structures, respectively. We also present guidelines to avoid the benchmarking pitfalls, so as to obtain accurate and representative performance measurements.  We hope that our work raises awareness about and provides a deeper understanding of some benchmarking pitfalls, and that it paves the way for a more rigorous, fair, and reproducible benchmarking.
\newpage